\documentclass[]{article}

\usepackage{amssymb,amsmath}
\usepackage{graphicx,color}
\usepackage{graphics,times}

\def\ve {\varepsilon}
\newtheorem{remark}{Remark}

\usepackage[colorlinks]{hyperref}
\hypersetup{
    linkcolor=blue,
}

\title{A Model for Collective Dynamics in Ant Raids }

\usepackage{authblk}

\title{A Model for Collective Dynamics in Ant Raids}

\author[1]{Shawn D.~Ryan\thanks{sryan18@kent.edu, The work of SR was supported by National Science Foundation Grant DMS-1212046.}}

  \affil[1]{Department of Mathematical Sciences and Liquid Crystal Institute\\
    Kent State University,  Kent, OH 44242 USA\\ }

\begin{document}

\maketitle

 
 \begin{abstract}
Ant raiding, the process of identifying and returning food to the nest or bivouac, is a fascinating example of collective motion in nature.  During such raids ants lay pheromones to form trails for others to find a food source.  In this work a coupled PDE/ODE model is introduced to study ant dynamics and pheromone concentration.  The key idea is the introduction of two forms of ant dynamics: foraging and returning, each governed by different environmental and social cues.  The model accounts for all aspects of the raiding cycle including local collisional interactions, the laying of pheromone along a trail, and the transition from one class of ants to another.  Through analysis of an order parameter measuring the orientational order in the system, the model shows self-organization into a collective state consisting of lanes of ants moving in opposite directions as well as the transition back to the individual state once the food source is depleted matching prior experimental results.  This indicates that in the absence of direct communication ants naturally form an efficient method for transporting food to the nest/bivouac.  The model exhibits a continuous kinetic phase transition in the order parameter as a function of certain system parameters.  The associated critical exponents are found, shedding light on the behavior of the system near the transition.  

 \end{abstract}

 \section{Introduction}

Collective motion in active biological systems has been of significant recent interest from flocking birds to collective swimming of microorganisms (e.g., \cite{Ari14,Rya11,Rya13a,Rya13b,Ari11,Sum12}).  Though this emergent behavior has been extensively studied, it has yet to be fully understood.     Collective motion offers many advantages  over individual motion to a given population of organisms including improved mixing, increased diffusion, faster transport, and new effective properties.  The main focus of works over the past few decades has been the study of bacterial suspensions with fewer focusing on insect swarming.  This emphasis on collective swimming of microorganisms is mostly due to the extensive amount of available experimental data (for a review see \cite{Cam97}).  However, social behavior in insects is the first form of collective motion one encounters in early childhood.  For example, the migration of butterflies and moths during season changes,  swarming bees, and trail formation by ants while foraging for food referred to as {\it raiding}. The main feature common to all active biological systems, in contrast to classical passive systems,  is the presence of self-propelled motion.   For a thorough review of past works on general animal populations, see \cite{Vic12}, and for a review of recent works on active biosuspensions, see \cite{Ara13,She13}.

We now briefly provide a review of the general behavior of ants extracted from the detailed experimental observations in \cite{Got95,Hol90,Sch71,Sch40}.   Raiding is common to all ants such as fire ants {\it Solenopsis invicta} \cite{Tsc06}, but we focus specifically on  army ants of the genus {\it Eciton}  (e.g., {\it Eciton hamatum} or {\it Eciton burchelli}) \cite{Sch71}.   A single army ant colony can consist of up to 200,000 ants and transport up to 3000 food items per hour up to 100m \cite{Cou03}.  The general cycle of life for a particular colony consists of bivouac formation (nest composed of living ant bodies), raiding, and migration.     A typical ant raid is carried out during the daylight hours to avoid predators and to allow for time in the evening for the colony to relocate the nest or bivouac under the cover of darkness \cite{Cou03,Sch40}.  Therefore, an efficient raiding process is crucial for the development and maintenance of a colony (a similar need for efficiency was investigated in the case of honeybees via a mathematical model introduced in \cite{Sch10}).


Raiding itself can have two forms: (i) {\it column raids} where ants form narrow bands of chemical pheromone trails to and from a food source and (ii) {\it swarm raids} where ants hunt as a large mass and move as essentially a single body \cite{Bat63}.   
 This work focuses on the column raids where, in the absence of direct communication, ants rely on the detection of pheromone trails laid by others to both find known food sources or to return to the nest once food has been found.  The ant raiding process consists of three main steps:
\begin{list}{}{}
\item{(i)} Initially some ants, referred to as foragers, leave the nest to perform essentially a random walk in search of food.
\item{(ii)} Once food is found ants lay a special chemical to mark the food location and continue to lay the chemical along the trail back to the nest to attract others.
 \item{(iii)} When an ant returns with food it, along with other foragers, begins to follow the newly created chemical concentration gradient back to the food.
 \end{list}
      This chemical gradient is composed of pheromones and has many additional purposes such as transmitting messages about predators or identifying one colony from another \cite{Sum03}.  This cycle continues thereby keeping the trail pheromones from dissipating until the food source has been depleted.   

This marking procedure leads to fascinating collective phenomenon including the formation of ``super highways" consisting of ants traveling back and forth forming lanes for increased mobility as recently observed experimentally in \cite{Cou03,Dus09,Fra85}.  This lane formation is similar to a group of people in a crowded crosswalk at a busy intersection.  To ensure everyone makes it to where they are going as efficiently as possible, unconsciously individuals form lanes for increased mobility.



While pheromones play a crucial role in raiding, other local interactions are also important in the dynamics such as collisions.   To truly understand the raiding behavior one must develop a model capable of investigating the effects of the relevant physical parameters such as ants size, chemical concentration, receptiveness to pheromone, and noise in foraging on the emergence of a collective state.  While there are countless biological studies on ant behavior, only recently has mathematics been used to further understanding.  Various recent mathematical approaches to modeling and simulation have been capable of capturing remarkable results such as lane formation \cite{Cou03}, pheromone trails resulting from collective behavior \cite{Deg13}, and the emergence and depletion of trails based on the concentration of a food source \cite{Amo14} among others (e.g., \cite{Joh06,Sch97,Wat95a,Wat95b}).  Specifically, the efficiency in which ants form and follow trails as well as the self-organization of a colony into a collective state has been examined experimentally in \cite{Gar13,Gar09,Per12} and with an individual based model in \cite{Vit06}. 

 A recent continuum model presented in \cite{Amo14} provides interesting results on ant foraging exhibiting spontaneous trail formation and efficient food removal.   While continuum approaches offer the advantage in general of being computationally efficient, they lack the ability to study interactions at the microscopic level and their effect on the resulting macrostate as well as only offering results ``on average".      Other previous models seeking to capture trail formation, such as \cite{Cou03}, impose an artificial pheromone gradient or a directional preference in the ants from the onset without allowing for it to be produced by the system itself.

    
 In contrast, this work seeks to improve on past models and provide additional insight through the development of a new first principles coupled PDE/ODE model for the pheromone concentration and ant dynamics respectively using basic principles learned over time from the immense works on modeling bacterial suspensions (including our own \cite{Rya11,Rya13a,Rya13b}).    As in the study of bacterial suspensions we seek to strip the model of inessential features and leave only those, which truly account for collective behavior.  Our model allows for the direct investigation of individual interactions at the microscopic level and their contribution to both the onset of collective behavior as well as  local traffic lane formation. This work will show that lane formation naturally results from each ant's desire to avoid collisions, which impede their motion.  Also, a posteriori we observe the model has a further advantage in that it allows one to show a continuous kinetic phase transition with respect to certain physical parameters and investigate the critical behavior in the population of foragers near the transition.


Though previous approaches have been developed to study the ant raiding cycle based on a continuum PDE \cite{Amo14,Joh06,Wat95a}, to the author's knowledge this is the first coupled PDE/ODE model for the entire raiding cycle, which focuses on the movement of individuals rather than the density of ants. The main benefit of developing this sound mathematical theory is that experiments have limitations such as observation time and a lack of control over some parameters (e.g., diffusivity of pheromone or amount of pheromone deposited).  Analysis of the model will also lead to a better understanding of ant behavior, which will have many ecological impacts in both conservation and pest termination.  This paper adds to the current knowledge on the way to understanding even more complex biological systems such as birds, fish, and potentially even humans.  The purpose of this work is to introduce a new model for ant raiding and show two main results (i) the transition to the collective state and (ii) the formation of lanes for efficient transport of food back to the nest.  Both are investigated in the case of one and multiple food sources.

 

What separates this work from most others is that our model accounts for the entire raiding process starting from foraging for food, forming a trail, and the resulting transition to the collective raiding state.   In addition, the model presented herein will deal with the depletion of food and the resulting transition back to individual behavior.  In Section~\ref{sec:model}, the main assumptions governing ant raiding are introduced and the coupled PDE/ODE model for ant dynamics and pheromone concentration are developed.  Separate equations of motion are introduced for ant foragers and those returning to the nest with food.  In Section~\ref{sec:results}, results are presented showing a clear phase transition to collective motion through the course of raiding and a transition back to individual motion when the food is depleted.  Also, evidence of lane formation along the pheromone trail is presented illustrating the macroscopic traffic-like dynamics formed from the local microscopic interactions matching prior experiment \cite{Cou03}.  The model is then used to investigate collective dynamics in the case of multiple food sources revealing some differences than in the single raid case.  In Section~\ref{sec:pt},  continuous kinetic phase transitions in the order parameter are shown and the corresponding critical exponents are found.  This illustrates the behavior of the system near the transition to collective motion as a function of relevant biological parameters such as the strength of noise in the system, the rate of pheromone diffusion, and the amount of pheromone deposited.  Also, a connection is made to classical thermodynamic systems with similar critical behavior.  Finally in Section~\ref{sec:disc}, the results are discussed and related to current biological knowledge as well as outlining potential future additions to make the model even more robust.

 \section{Model}\label{sec:model}
 
The key idea behind the model developed herein is to divide the ants into two classes: foragers and returners each with different equations of motion, because each is motivated by different environmental and social cues  \cite{Bec92,Wil62}.    The similarity between the two classes is that both are self-propelled and want to avoid collisions with one another, yet they differ in their attraction to the chemical gradient or lack thereof. 
 
\begin{figure}
\centerline{\includegraphics[height=1.25in]{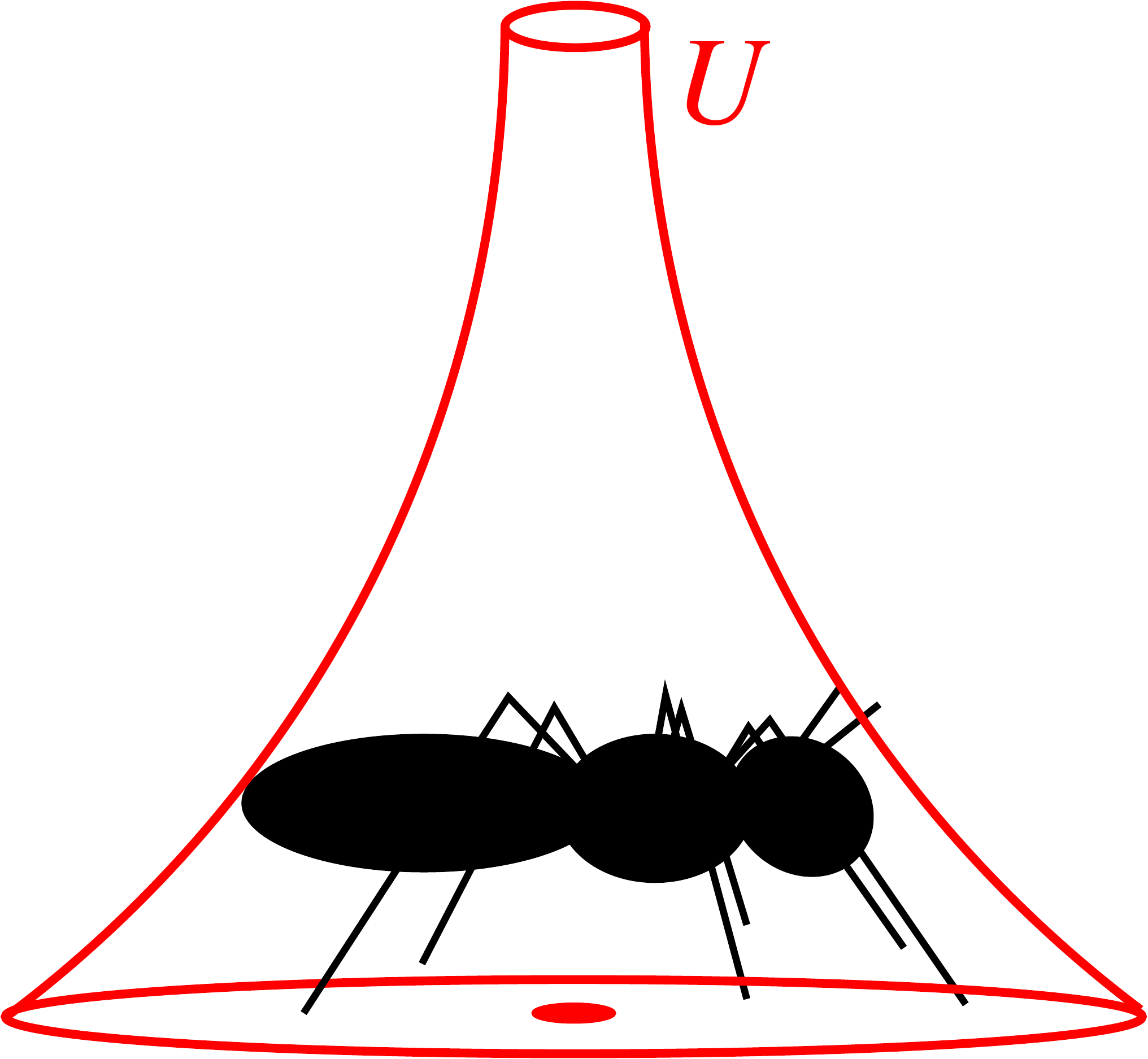} \hspace{.75in} \includegraphics[height = 1.25in]{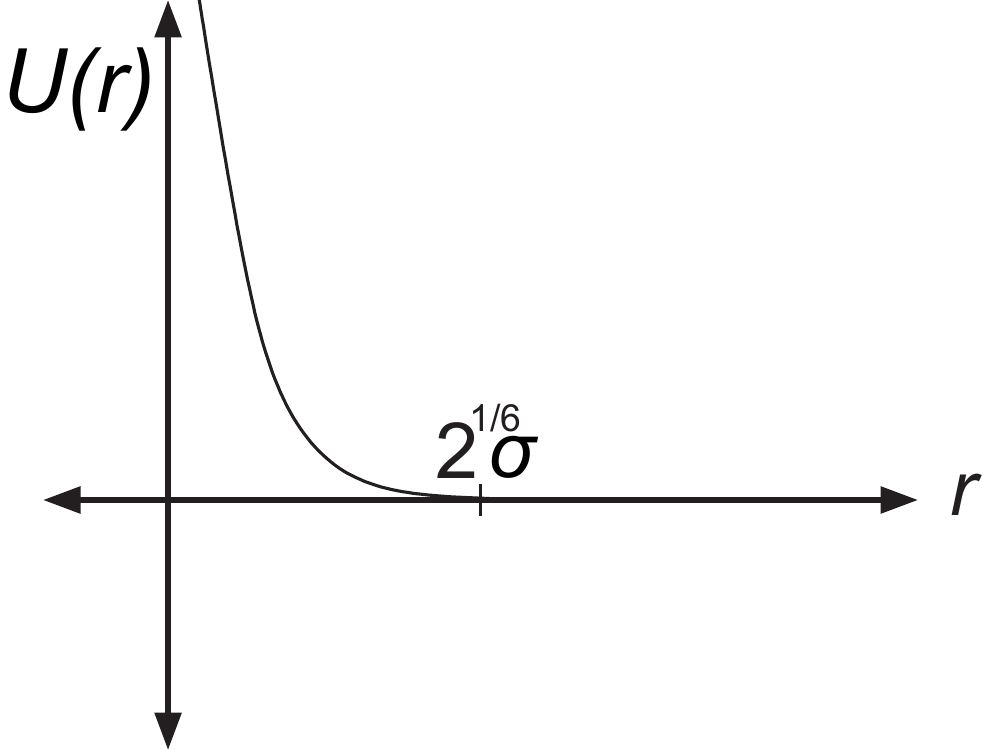}}
\caption{Left: Ant represented by a point particle with an excluded volume determined by the truncated repulsive potential $U$.  Right: Truncated Lennard-Jones type potential is radially symmetric with $r = |{\bf x}|$.  }\label{fig:repulsive}
\end{figure}

 We represent each individual ant as a point with an excluded volume, see Figure~\ref{fig:repulsive}.   The center of mass and velocity for an individual ant are governed by ODEs describing the evolution of each in time.  We suppress the details of the ant body  so that the simulation of such a model is made simpler, yet still captures the desired results.  To account for the correct behavior, it is crucial to model the pheromone diffusion carefully \cite{Deg13}.  Thus, a critical component in the equations of motion for each ant is the contribution from the pheromone gradient, denoted $\nabla c({\bf x},t)$, where the pheromone concentration $c({\bf x},t)$ satisfies a parabolic reaction-diffusion PDE introduced in Section~\ref{sec:pher}.

 \subsection{Assumptions}\label{sec:assump}

Ants in the foraging phase leave the nest location in search of food.  Until a food source is found no pheromone gradient exists and motion is dominated by a random walk \cite{Amo14,Deg13,Sch97}.  After a food source is identified and marked by pheromone, then  other foragers, which encounter the pheromone gradient, follow it.  After reaching the food source each ant becomes a returner and follows a direct path back towards the nest ignoring the pheromone concentration gradient.   When a returner reaches the nest or bivouac it transitions back to being a forager and the cycle repeats.  
 

It is assumed throughout this work that returners know where their home is and take the most direct path toward it.   The fact that the path back is direct has been observed experimentally in \cite{Bue14,MuhWeh88,Nar13,Weh03} where even detouring ants by imposing barriers after the food source is found does not dissuade them from following the most direct path. In these works it is noted that ants can follow landmark routes and recognize locations to navigate.  Their evidence suggests that ants can use path integration and their knowledge of complex outbound routes  to return home along a straight path.  Ants do not use complicated path integration in the same way as a human, but rather use an approximation accounting for navigational errors \cite{MuhWeh88}.

 We assume the colony of ants are self-propelled particles represented by a set of points $\{{\bf x}_i\}, i = 1, ..., N$.  Each point can be thought of as the location of the center of mass for an individual ant.  The velocities of each ant $\{{\bf v}_i = \dot{\bf x}_i\}, i = 1, ..., N$ are tracked as well as their orientation ${\boldsymbol \omega}^i := {\bf v}^i/|{\bf v}^i|$.  Even though an ant is modeled as a point, in reality, the centers of mass for two ants cannot become arbitrarily close due to the presence of the body.  Instead of having to resolve the details of finite size body interactions, which can be computationally expensive,  we introduce a truncated repulsive Lennard-Jones (LJ) type potential $U(|{\bf x}|)$ as a function of the interparticle distance $r = |{\bf x}|$
\begin{equation}\label{eqn:truncated}
 U(r) = \begin{cases} 4\ve\left[\left(\frac{\sigma}{r}\right)^{12} - \left(\frac{\sigma}{r}\right)^6\right] + \ve, & \quad r \leq 2^{1/6}\sigma,\\
 0, & \quad r > 2^{1/6}\sigma
 \end{cases}
\end{equation} 
 where $\ve$ represents the strength of repulsion.  The effective size of an ant is defined by the length $\ell = 2^{1/6}\sigma$ where the repulsive forces between two ants balance to zero.  In principle each ant $i$ can interact with any other $j$, but these interactions are local in nature.  This force from the $j$th ant on the $i$th ant is defined as ${\bf F}({\bf x}_i - {\bf x}_j) := -\nabla_{\bf x} U({\bf x}_i-{\bf x}_j)$ depending only on the relative distance $r = |{\bf x}_i-{\bf x}_j|$.   By introducing the potential \eqref{eqn:truncated}, the collisions between individual ants are modeled as a soft excluded volume interaction.  The truncated potential has also been successfully used in our recent work on bacteria to impose an effective size on a point particle \cite{Rya11,Rya13a,Rya13b}.      Now one must account for an ant's response to the local pheromone concentration.  

\begin{remark} 
Ants have elongated bodies and its possible to incorporate this shape through a truncated elliptical potential (e.g., a modified Gay-Berne potential \cite{Qi12}), but is not needed to achieve the desired results and introduces a greater computational expense since the resulting force has an added dependence on the orientation of each ant.
\end{remark}

 \subsection{Pheromone concentration}\label{sec:pher}
 
When a foraging ant detects the chemical signature of the pheromone it will use its antennae to analyze the local concentration and decide which direction to travel \cite{Deg13,Cal92,Cou03}.  For an extensive review of the background of chemotaxis in ants and its implications for their movement consult \cite{Amo14} and references therein.  Pheromone deposition and trail laying are well modeled by a two-dimensional reaction-diffusion process for the chemical concentration $c({\bf x},t)$
\begin{equation}\label{eqn:pher}
\begin{cases}
\displaystyle\partial_t c - \alpha\Delta c + \gamma c = \sum_{j = 1}^M qe^{-\|{\bf x}_j(t)-{\bf x}_f\|^2}\delta({\bf x}-{\bf x}_j(t)), &  {\bf x} \in \mathbb{R}^2, t >0\\
c({\bf x}, 0) = g({\bf x}), &  {\bf x} \in \mathbb{R}^2.
\end{cases}
\end{equation}
Here ${\bf x}_f$ is the location of the food with initially $M$ food items, $\alpha$ is the diffusion coefficient controlling the rate at which the pheromone spreads, and $\gamma$ is the evaporation coefficient that ensures an exponential decay of the pheromone in time.  The coefficient $qe^{-\|{\bf x}_j(t)-{\bf x}_f\|^2}$ represents the amount of pheromone deposited at time $t$ and decays as a returning ant moves away from the food source.  This decrease is needed to ensure that the proper gradient forms due to the competition with diffusion.   The function $g({\bf x})$ represents the initial distribution of chemical, which is taken as uniform, $g({\bf x}) = const$, or zero so there is no pre-defined directional preference.

The coefficient $\gamma$ plays an important role in raiding, because trails that no longer lead to viable food sources should be removed quickly for maximum efficiency.  The trail is defined as the line segment connecting the food source ${\bf x}_f$ to the nest denoted by location ${\bf x}_c$. In this work we can see the trail naturally form by studying the transition to the collective state and the deviation of individuals from the trail center.    

Equation \eqref{eqn:pher} captures the exponential decay of the concentration as well as the diffusion to the surrounding environment.    The coupling of the PDE for pheromone concentration \eqref{eqn:pher} to the ODEs governing ant dynamics introduced in Section~\ref{sec:eom}  is analogous to PDEs for chemotaxis (such as Keller-Segel \cite{Kel70,Kel71}), which have been used prominently in models for swimming microorganisms \cite{She12,Xue13}. 

 \subsection{Equations of motion}\label{sec:eom}
 
 We now derive the equations of motion for the evolution of the particle centers of mass $\{{\bf x}_i\}_{i=1}^N$ from a balance of forces.  The two distinct dynamic models for foraging and returning ants respectively are composed of the different forces associated to each group's behavior.
 
\subsubsection{Foraging ants}  

Before the food source is identified, foraging ants perform a random walk, propel themselves in the direction they are currently oriented and try to avoid collisions with other ants.  Once a food source has been discovered the location,  ${\bf x}_f$, is marked and pheromone starts to diffuse into the surrounding environment.  As previously discussed the collisions will be modeled via a short-range repelling potential described in Section~\ref{sec:assump} and the pheromone gradient will be induced by a solution to the reaction-diffusion equation \eqref{eqn:pher} presented in Section~\ref{sec:pher}.

To make the physical description complete we now introduce the ODE model for the dynamics of foraging ants
\begin{equation}\label{eqn:ibm-f}
\begin{cases}
\dot{\bf x}_i &= {\bf v}_i\\
\displaystyle\dot{\bf v}_i &=  \nu{\bf v}_i\left(\xi^2 - |{\bf v}_i^2|\right) - \frac{1}{N}\sum_{j \neq i} \nabla_{\bf x} U(|{\bf x}_i - {\bf x}_j|) + d\nabla_{\bf x}c({\bf x},t)+DW_t
\end{cases}
\end{equation}
where $U$ is a repulsive potential \eqref{eqn:truncated}.  The force of self-propulsion is proportional to the velocity via the coefficient $\nu\left(\xi^2 - |{\bf v}_i^2|\right)$.  Observe that the term $\left(\xi^2 - |{\bf v}_i^2|\right)$ ensures exponential growth or decay to the isolated translational speed $\xi$.

The pheromone concentration, $c({\bf x},t)$, enters with relative strength $d > 0$ representing the sensitivity of the ants to the chemical gradient when present.  The random walk is controlled through the strength of the noise $D$ and a Gaussian white noise process $W_t$ with mean zero and variance one.  This white noise process has two purposes: (i) to enforce the foraging behavior as a random walk and (ii) it can represent a level of misinformation in detecting the chemical trail or a lack of receptivity to the chemical stimulus.  A similar approach incorporating stochastic terms has been used recently in \cite{Bur12}   for studying general aggregation of individuals and \cite{Erb12,Esc10} for the behavior of locusts.  These dynamic equations are coupled to the PDE \eqref{eqn:pher} introduced in Section~\ref{sec:pher}.   

These equations contain three competing factors controlling individual ant dynamics: (i) self-propulsion, $\nu{\bf v}_i\left(\xi^2 - |{\bf v}_i^2|\right)$, (ii) excluded volume / collisions, $-\nabla_{\bf x} U$, and (iii) pheromone concentration gradient, $d\nabla_{\bf x} c$.  The interplay between these three forces leads to the transition from individual to collective behavior.

\begin{remark}
The truncation and vertical translation of the original Lennard-Jones 6-12 potential (see Figure~\ref{fig:repulsive}) was imposed so that the force ${\bf F} = -\nabla U$ would be Lipschitz continuous.  This will prove important if one wants to show the longtime existence for the ODE particle equations of motion (previously done for bacteria in \cite{Rya13a}).
\end{remark}
 
 \subsubsection{Returning ants}

Once a foraging ant comes into contact with a food source it becomes a returning ant.  After acquiring food, the ant proceeds to take the minimal path back to the nest, which is assumed to be a straight line ignoring environmental effects such as elevation or obstacles.  
 As an ant journeys home it still propels itself in the direction it is oriented and tries to avoid collisions with others.  To make the physical description complete we now introduce the ODE model for the dynamics of returning ants
\begin{equation}\label{eqn:ibm-r}
\begin{cases}
\dot{\bf x}_i &= {\bf v}_i\\
\displaystyle\dot{\bf v}_i &= \nu{\bf v}_i\left(\xi^2 - |{\bf v}_i^2|\right) - \frac{1}{N}\sum_{j \neq i} \nabla_{\bf x} U(|{\bf x}_i - {\bf x}_j|) + \beta\frac{{\bf x}_i - {\bf x}_c}{r}
\end{cases}
\end{equation}
where $({\bf x}_i - {\bf x}_c)/r$ is the unit vector directed to the nest with $r = |{\bf x} - {\bf x}_c|$.  The coefficient $\beta$ governs the relative strength of an ant's desire to return to the nest.  As in the dynamic equations for foraging ants, the self-propulsion is represented by $\nu{\bf v}_i\left(\xi^2 - |{\bf v}_i^2|\right)$ and the truncated repulsive potential is $U$.   

The equations of motion \eqref{eqn:ibm-f}-\eqref{eqn:ibm-r} have a similar form to those developed in \cite{CarForTosVec10} and are reminiscent of {\it D'Orsogna et al.}  \cite{Ber06} who considered the stability of collective structures and milling of particles with a similar individual based model (IBM).  The coupled PDE/ODE model developed in this work provides a more realistic description of the movement, trail laying, and interaction at the microscopic level as compared to previous ODE models for ants restricted to a lattice (e.g., \cite{Sol00}).   For the non-dimensionalization of the system and values for the relevant biological parameters see Appendix~\ref{app:nd}.  Before providing the details of the numerical implementation of the model, we introduce the order parameter used to measure the correlated behavior of the system.   

\subsection{Order parameter}\label{sec:op}

In order to quantify how correlated the particles are in the system, we introduce a reasonable order parameter referred to as the {\it flow}, $F$,
\begin{equation}\label{eqn:op}
\displaystyle F = \frac{1}{N} \left| \sum\limits_{i = 1}^N {\boldsymbol \omega}_i \right|
\end{equation}
where $N$ is the number of ants under consideration and ${\boldsymbol \omega}_i = \frac{{\bf v}_i}{|{\bf v}_i|}\in \mathcal{S}^1$ represents each ants orientation.   If each individual ant moves in an arbitrary direction, the velocity vectors will effectively cancel each other giving a flow of $F = 0$ representing a disordered phase.  If all the ants move in the same direction (i.e., toward a food source or the nest), then $F \approx 1$ representing an ordered phase.  

With this order parameter we can investigate the phase transition that occurs during the course of an ant raid, but some care must be taken in how to apply this definition.  For instance, during the course of the raid the ants will form trails of incoming and outgoing ants moving in opposite directions.  In terms of the order parameter, $F$, these two groups would effectively cancel each other's contributions resulting in a net flow near zero.  

Since we naturally consider two types of ants, foragers and returners, each governed by different dynamic equations, we must consider their flows separately.  Thus, the entire system will be described by two order parameters: $F_{for}$ and $F_{ret}$ using definition \eqref{eqn:op}, but only summing over the relevant ants.  Even though the ants change from one group to another numerous times through the course of a raid we will observe that each population still exhibits collective behavior when considered separately.  This particular choice of order parameter for systems of self-propelled particles was first utilized to the author's knowledge by {\it Vicsek et al.} in \cite{Vic95} and applied more recently to ants in \cite{Cou03}.  Throughout Section~\ref{sec:results} this order parameter will be used to investigate the effects of biophysical parameters present in the model on the collective state and in Section~\ref{sec:pt} it will allow one to show a continuous kinetic phase transition as a function of the those parameters.

\subsection{Numerical implementation}\label{sec:num}

 Numerical implementation of the coupled PDE/ODE model \eqref{eqn:pher}-\eqref{eqn:ibm-r} is rather straightforward due to its simple nature.  Here we merely highlight a few of the more interesting points that need to be considered when carrying out the simulations.  One of the advantages of this model is the fact that we can write down an explicit solution to \eqref{eqn:pher}.  Assuming an initial uniform distribution $g({\bf x}) = \frac{1}{|V_L|}$,
  we have the following expression for the pheromone concentration and its gradient
\begin{align}
\text{\small{$c({\bf x},t) := e^{-\gamma t}\left[\frac{1}{|V_L|} + q\left(\sum_{j = 1}^M\int_{t_{dis}}^t e^{-\|{\bf x}_j(s)-{\bf x}_f\|^2}\frac{e^{\gamma s}}{4\pi\alpha(t-s)}e^{-\frac{\|{\bf x}-{\bf x}_j(s)\|^2}{4\alpha(t-s)}} ds \right)\right]$}}\nonumber\\
\text{\small{$\frac{\partial c}{\partial x_i} = qe^{-\gamma t}\left[\sum_{j = 1}^M\int_{t_{dis}}^te^{-\|{\bf x}_j(s)-{\bf x}_f\|^2}\frac{-2(x_i - x_i^{j}(s))e^{\gamma s}}{16\pi\alpha^2(t-s)^2}e^{-\frac{\|{\bf x}-{\bf x}_j(s)\|^2}{4\alpha(t-s)}}ds\right]$}}.\label{eqn:fund}
\end{align}
   The $i$th component of the returning ant at time $t$, ${\bf x}_j(t)$, is denoted $x_i^{j}(t)$.  This solution can be derived by using the fundamental solution to the heat equation
 \begin{equation*}
 \Phi({\bf x},t) := \begin{cases} \frac{1}{4\pi\alpha t}\text{exp}\left(-\frac{|{\bf x}|^2}{4\alpha t}\right), &\quad {\bf x} \in \mathbb{R}^2, t > t_{dis}\\
 0, & \quad {\bf x}\in\mathbb{R}^2, t<t_{dis}
 \end{cases}
 \end{equation*}
 and the relation $u({\bf x},t) = e^{\gamma t}c({\bf x},t)$ where $u({\bf x},t)$ solves the heat equation if and only if  $c({\bf x},t)$ solves \eqref{eqn:pher} (resulting in the so-called {\it Bessel potential}).  Thus, no finite difference approximation in space is needed when simulating the system.  We only need to impose a numerical integration technique such as a composite trapezoid rule to evaluate the time integral in \eqref{eqn:fund}.  This is the most time consuming part of the simulations, because it must be computed for each foraging ant.

\begin{table}
\caption{Values used in simulation for each of the dimensionless biological parameters.  See Appendix~\ref{app:nd} for biological values from prior experiments.}
\label{tab1}
\begin{tabular}{|l|l|l|}
\hline\noalign{\smallskip}
Parameter & Value & Physical Description\\
\hline\noalign{\smallskip}
$\nu$ & 1.0 & Strength of Self-propulsion\\
$\ve$ & .0001 & Strength of Repulsion Potential\\
$\sigma$ & .5 & Effective Ant length\\
$d$ & 10.0 & Pheromone Receptivity Strength\\
$\alpha$ & 10.0 & Pheromone Diffusion Coefficient\\
$\gamma$ & .001 & Pheromone Degradation Coefficient\\
$D$ & 1.0 & Strength of Noise in Random Walk\\
$\beta$ & 1.0 & Strength of Stimulus to Return to Nest\\
$q$ & 1.0 &	Amount of Pheromone Deposited\\
\noalign{\smallskip}\hline
\end{tabular}
\end{table}

To evolve the system in time a standard Forward Euler method is used
\begin{align*}
&\text{{\small  ${\bf x}_i(t + \Delta t) = {\bf x}_i(t) + {\bf v}_i(t+\Delta t)\Delta t$}}\\
&\text{{\small  ${\bf v}_i(t + \Delta t) = {\bf v}_i(t) + \Delta t\left[\nu{\bf v}_i\left(\xi^2 - |{\bf v}_i^2|\right) - \frac{1}{N}\sum\limits_{j \neq i}\nabla_{\bf x} U(|{\bf x}_i-{\bf x}_j|) - d\nabla_{\bf x} c({\bf x}_i)+DW_{t}\right].$}}
\end{align*}
The random walk implemented in the equations of motion is modeled via a discrete Gaussian white noise process $W_{t_{n+1}} = W_{t_n} + \sqrt{dt} \xi_{n+1}$ where $\xi_i$ is an i.i.d. Gaussian distributed random variable with mean zero and variance 1.  

 The basic computational domain can be arbitrary, but for the results presented it consists of a two-dimensional rectangle of non-dimensional length 100 $\times$ 50 allowing for trails around 200 times the size of an individual ant.  Reflecting boundary conditions are imposed so that the concentration of ants is conserved.  There are three cases of ants possibly hitting walls: i) An ant hits the top/bottom wall, then ${\bf v} = (v_x, v_y)$ is replaced by ${\bf v}_{new} = (v_x, -v_y)$, ii) An ant hits the left/right wall, then ${\bf v}_{new} = (-v_x, v_y)$, iii) An ant hits two or more walls (e.g., a corner), then ${\bf v}_{new} = (-v_x, -v_y)$.  However, once the trail begins to form and the collective state is reached the ants rarely reach the boundaries of the computational domain.  
 
 The typical time step for the simulations in dimensional form is $dt = .02s$ representing the temporal resolution of ants from experimental data \cite{Cou03}.  Typical simulations run for between 700,000 to 1,200,000 time steps, which translates to a typical ant raid of 4-7 hours consistent with observations of ants in nature from \cite{Sch71,Sch40}.  In this time between 2000-3500 food items are returned to the nest from the food source before it is depleted.  The simulations were run with random initial conditions and the results are averaged over numerous simulations.    The typical values of the non-dimensional parameters used in the simulations are given in Table~\ref{tab1} and for the relevant biological quantities see Appendix~\ref{app:nd}.

   \begin{figure}
 \centerline{\includegraphics[height=2.25in]{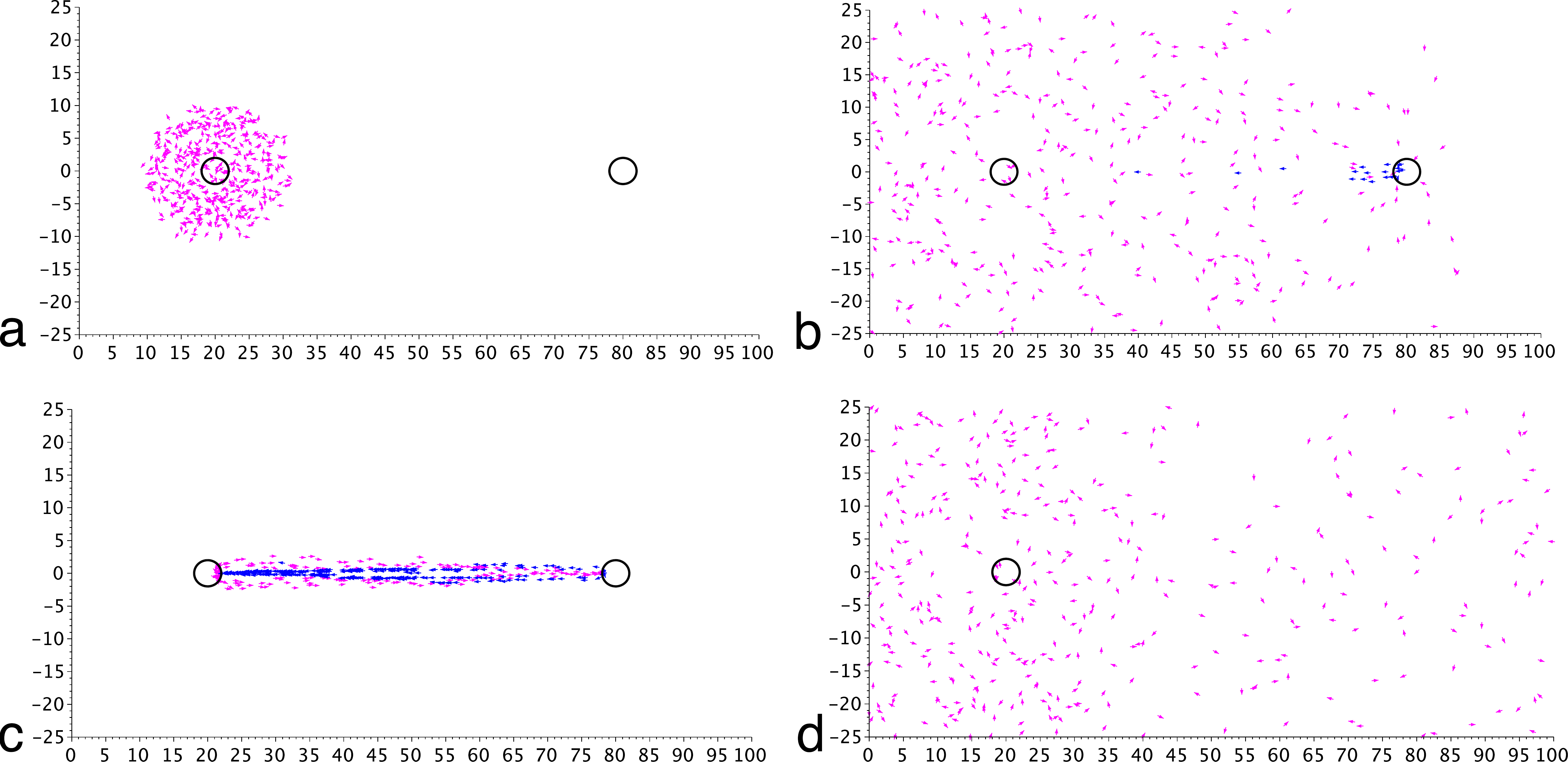}}
 \caption{Sample ant raiding simulations with foragers (purple) and returners (blue) where $N = 400$. Each arrow represents an individual ant's orientation ${\boldsymbol \omega}_i$.  The black circles denote the nest, ${\bf x}_c = (20,0)$, and food source, ${\bf x}_f = (80,0)$.  a) Initially ants are placed near the nest in non-overlapping positions with random orientation representing radial expansion outward from the nest. b) One of the foragers discovers the food source and marks it with pheromone, becoming a returner. As the pheromone diffuses more and more foragers detect the scent and begin to follow the trail to the food source. c) The trail forms displaying lanes of unidirectional flow. d) Once the food source is depleted the trail quickly disappears and the ants return to random foraging. See Online Resource 1.}\label{fig:trail}
\end{figure}
 
 \section{Results}~\label{sec:results}

\begin{figure}
 \centerline{\includegraphics[height=1.65in]{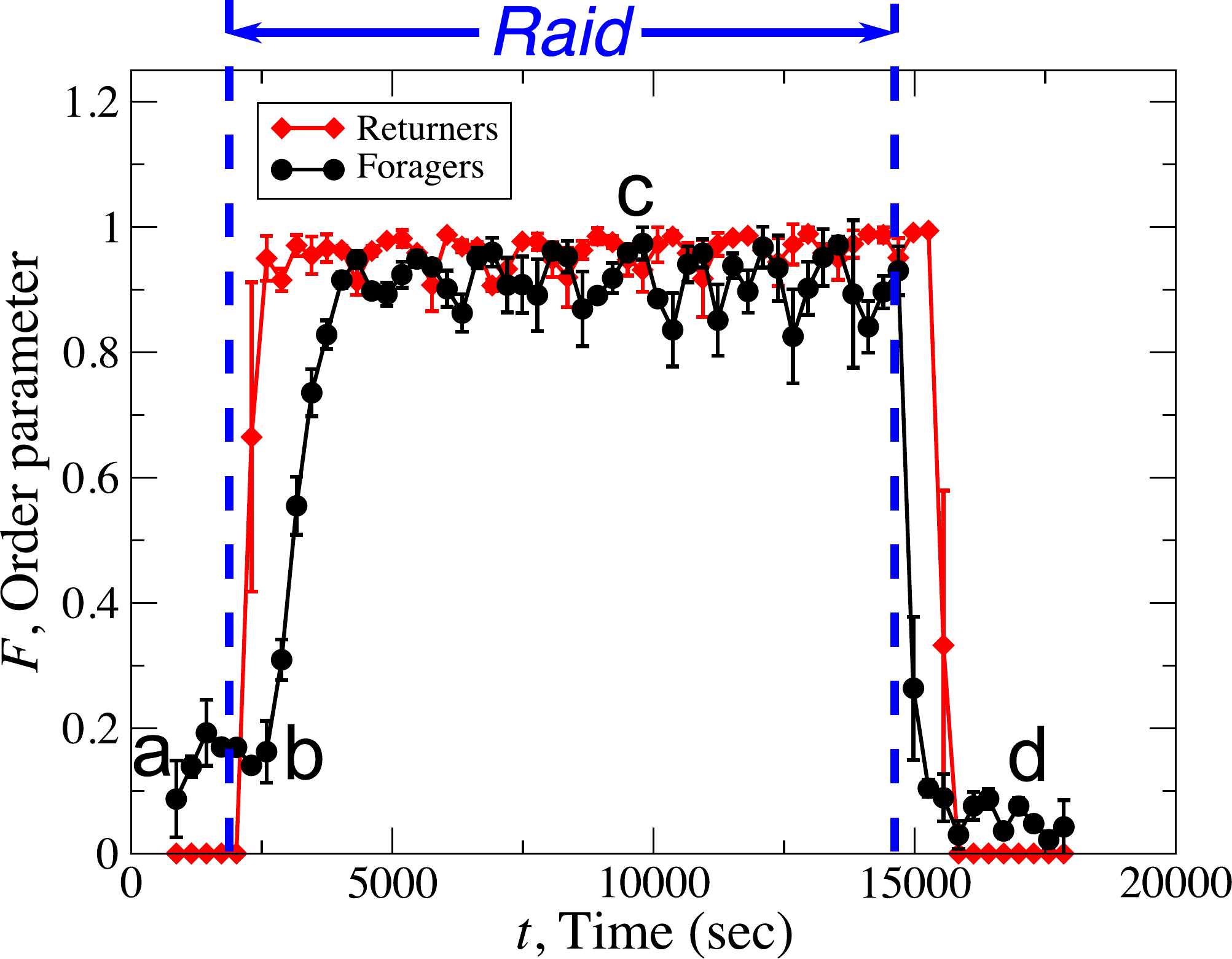}\hspace{.1in}\includegraphics[height=1.5in]{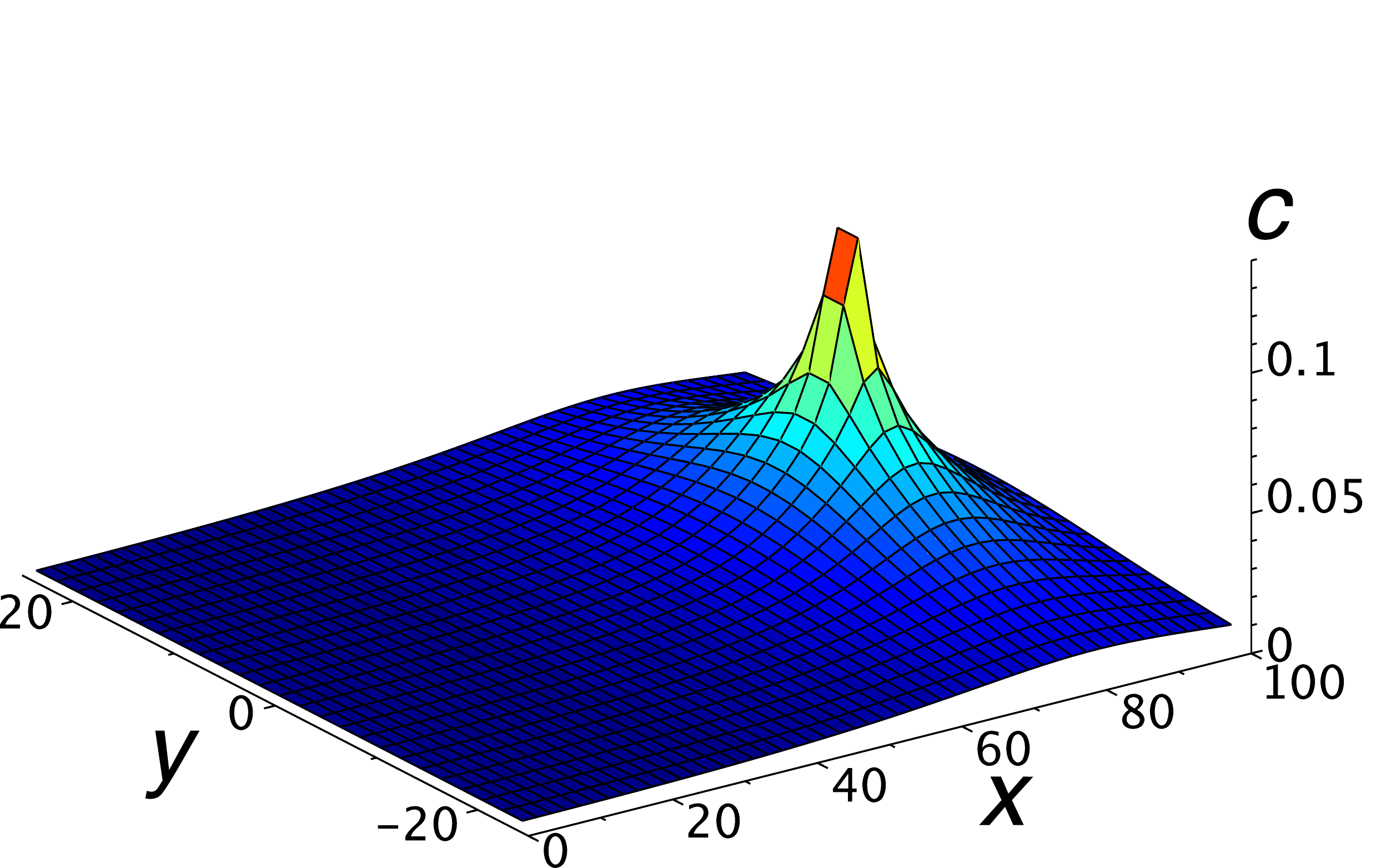}}
 \caption{Left: Order parameter $F = \frac{1}{N}\left|\sum_{i=1}^N {\boldsymbol \omega}_i\right|$ versus time showing a transition from the individual to the collective phase.  The time period of a raid is indicated in blue beginning at $t_{dis}$, the time of food discovery, and ending at $t_{dep}$, the time of food depletion.  Error bars represent one standard deviation.  The letters a)-d) correspond to the snapshots of the raid in Figure~\ref{fig:trail} for a typical simulation. Right: Typical profile for the phermone concentration $c({\bf x},t)$ once the trail has been established and the collecitve state has been reached. The peak in $c$ occurs at the location of the food source and exponentially decays away from the trail.  See Online Resource 2 for the evolution of the chemical concentration in time.}\label{fig:trans}
\end{figure}

\begin{figure}
 \centerline{\includegraphics[height=.9in]{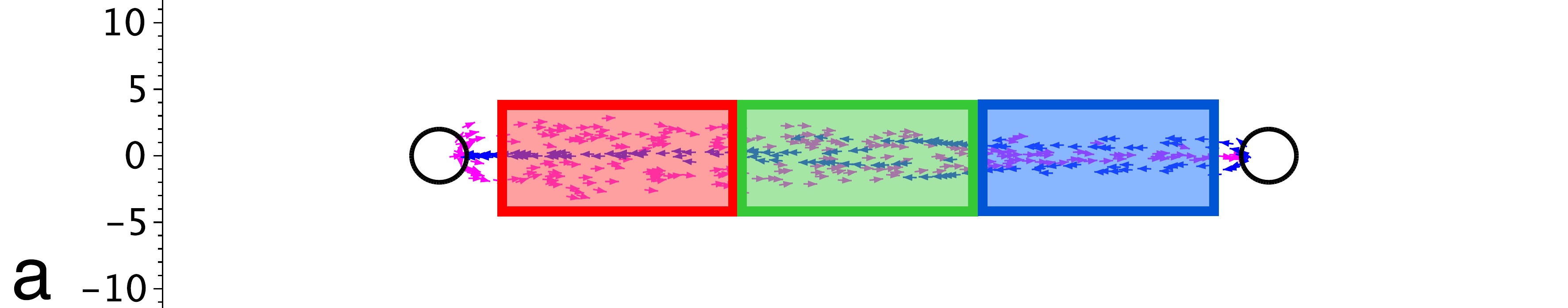}}

 \vspace{.075in}
 
  \centerline{\includegraphics[height=1.15in]{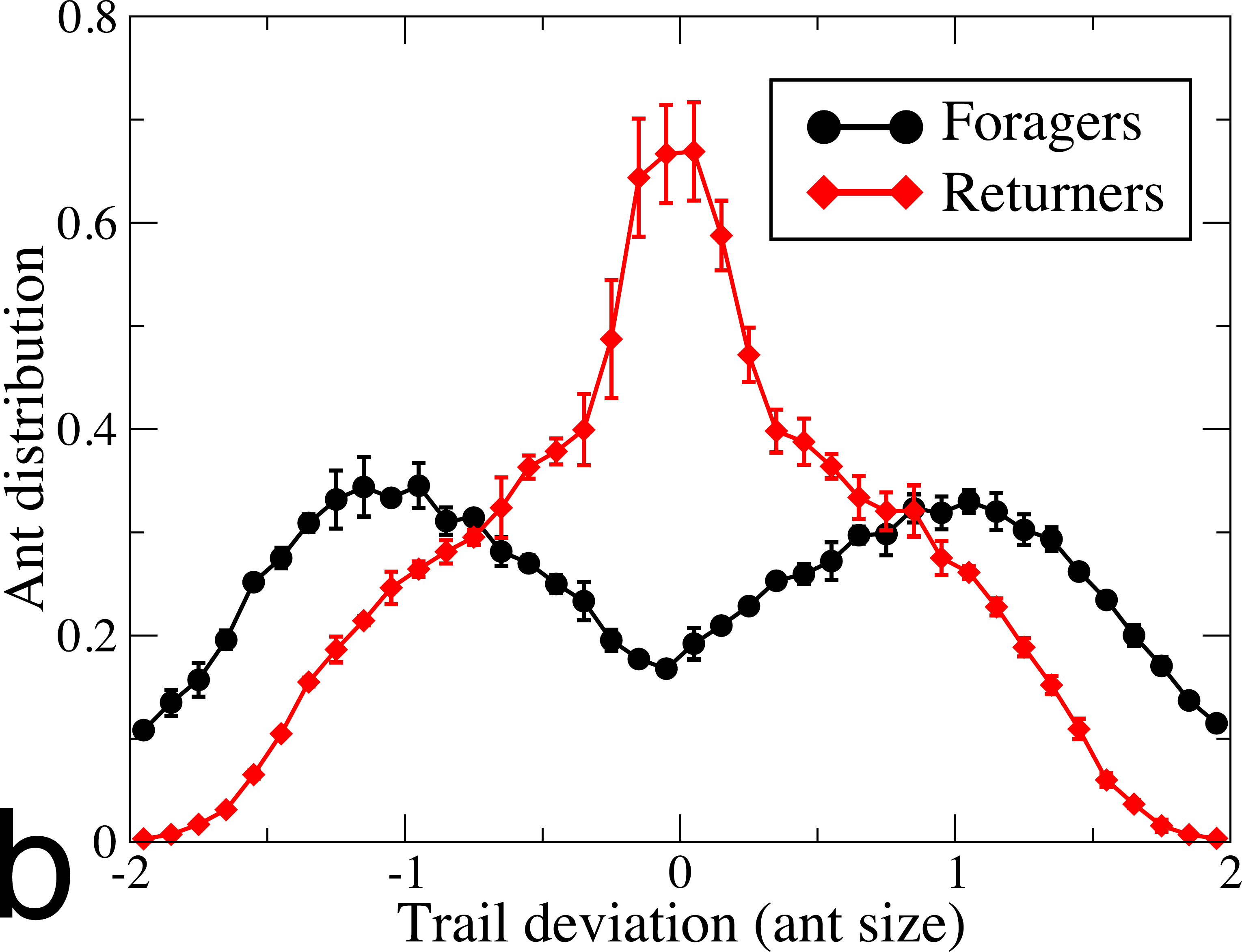}\hspace{.05in}\includegraphics[height=1.15in]{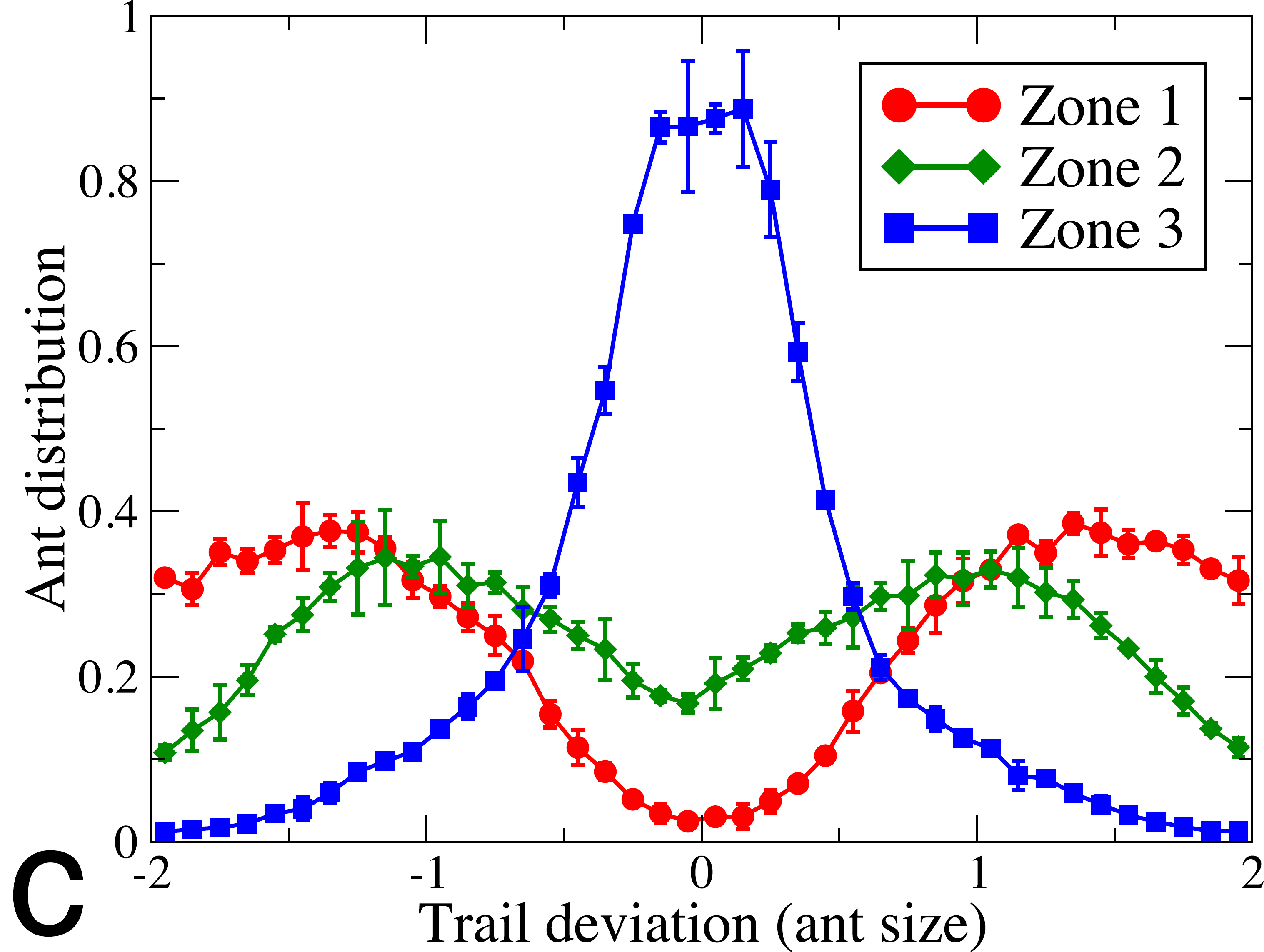}\hspace{.05in}\includegraphics[height=1.15in]{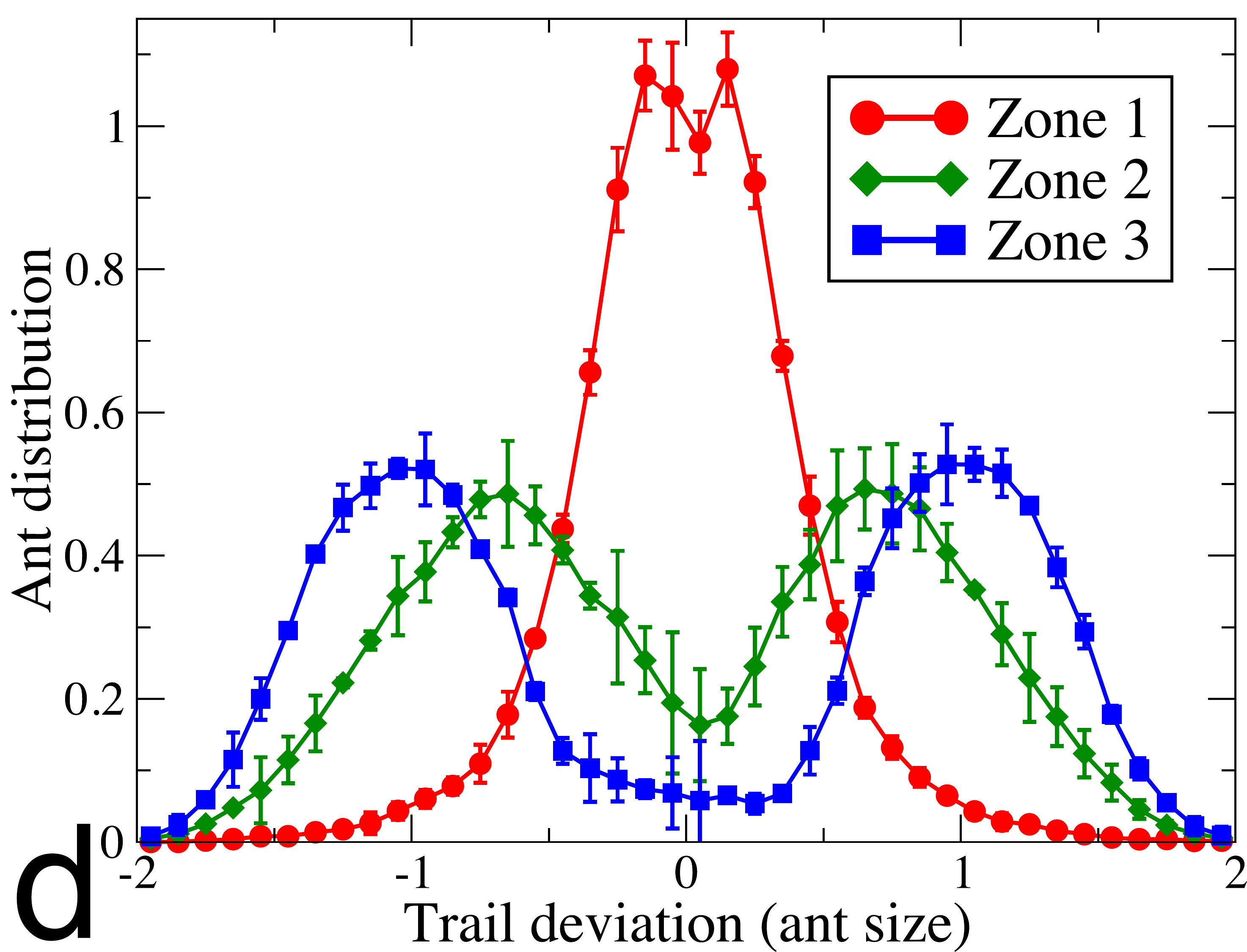}}
 \caption{Formation of lanes along the pheromone trail for foragers and returners.  Distances normalized by ant size $\ell=x_0$ (characteristic length).  a) The trail is broken into three distinct segments with different distributions of ants. b) The average ant distribution over the entire trail shows bi-modal peaks occur at a distance between $.5\ell-1.5\ell$ from the center indicating two outside lanes of foragers (black) with returners (red) in the middle.    The distribution of ants in each of the colored zones over time is found for c) foragers and d) returners.  Error bars represent one standard deviation.}\label{fig:lanes}
 \end{figure}

 In this section, we use numerical simulations as evidence that the model captures the swarming behavior found in army ant raids.  The two main results, which are evident from the Online Resources  and Figures~\ref{fig:trail}-\ref{fig:lanes} are (i) the transition of the system to collective behavior over time and (ii) the formation of lanes along the trail.   Specifically, one can see in Figure~\ref{fig:trail}a) that the ants start in a disordered state where each individual is randomly foraging for food until one ant finds a food source at time $t = t_{dis}$ (see Figure~\ref{fig:trail}b)).  The circular initial configuration is similar to that of a bivouac \cite{Sch71,Sch40}.  Once the food source is marked with pheromone the ants who have reached it begin returning to the nest (blue) while laying pheromone as nearby foragers begin to detect the increased chemical concentration (see Figure~\ref{fig:trail}c)).  Shortly after single lanes of ants begin to form.  This collective state is observed until food depletion at time $t = t_{dep}$ when random foraging resumes (see Figure~\ref{fig:trail}d).  A typical concentration profile for the pheromone once the trail has formed can be seen in Figure~\ref{fig:trans}.  For additional results starting from a centrally located nest see Appendix~\ref{app:center}.

 \subsection{Transition to a collective state}

 By introducing an order parameter \eqref{eqn:op} in Section~\ref{sec:op} that measures the coordinated behavior of each group we can study the transition to the collective state in time.  Figure~\ref{fig:trans} shows a sharp transition to collective behavior after the time of food discovery $t > t_{dis}$.  We notice that there is a time delay in the formation of a collective state of foragers due to the time elapsed from marking the food with pheromone and that pheromone diffusing out into the environment to be detected by others.  This time delay is due to the interplay between the diffusion term, $\alpha \Delta c$, and the term governing the exponential decay, $\gamma c$ in \eqref{eqn:pher}.  Figure~\ref{fig:trail}b) illustrates that locally near the food source where the pheromone has begun to diffuse the ants become attracted to the location of the food.  As the pheromone diffuses out to the whole domain and the trail is laid more and more foragers become attracted.  This can be seen by the steady  increase of the order parameter $F$ for the foragers.  While some may argue the returning ants reach a collective state in a trivial way due to the fact that the go directly to the nest, this does not account for the lane formation that will be discussed further in Section~\ref{sec:lane}.

 The raiding trail is considered to be formed when the order parameter for both the returners and the foragers is near one as illustrated in Figure~\ref{fig:trans}).  Once the food is depleted at $t = t_{dep}$ we observe a rapid decrease of the order parameter.  This is due to the fact that ants no longer lay chemical at that location and the pheromone evaporates exponentially fast.  Since the chemical gradient has no bearing on the returning ants, the foragers deviate from collective behavior first.  Those returning still must deliver the food they have to the nest  along the home-bound vector before going on to other functions.  After this time all the ants are foragers and in the absence of a chemical gradient or detection of a new food source all ants merely perform random walks returning to a disorder state.  

 \subsection{Lane formation}\label{sec:lane}
 
 In addition to a transition to and from the collective state, we can also consider the local behavior along the trail.  In particular, a histogram of the position of each ant with respect to its distance from the trail center is used to form an ant distribution function in the neighborhood of the trail.  As can be seen in Figure~\ref{fig:lanes}, the foragers and returners naturally self-organize into lanes like cars on a highway or people in a crosswalk.  Specifically, foragers who are driven by the chemical gradient occur between .5-1.5 ant lengths from either side of the trail with the highest probabilities forming a bi-modal distribution (see Figure~\ref{fig:lanes}b)).  Whereas, returners who are driven by their desire to return to the nest as quickly as possible occur at the trail center with highest probability.  We conclude, for the majority of the time along the trail, one lane of returners forms in the trail center and two lanes of foragers flank each side with equal probability.  
 
 The formation of three lanes is consistent along the whole trail, but which class of ant is in the middle varies.  By employing a microscopic model, unlike a continuum model, we can study different regions of the trail and focus on the local behavior (see Figure~\ref{fig:lanes}b)).  Near the nest returning ants are in the center and foraging ants leave on either side with equal probability (Zone 1, red).  In the central region (Zone 2, green) there is a crossover event where the populations switch lanes and foragers move toward the middle as they get closer to the food source.  Here even 5 or 7 lanes of alternating classes of ants can be observed if the density of ants is large compared to the trail length.  Near the food source (Zone 3, blue) the chemical gradient is strong and returners have equal probability of leaving with food on either side of the trail.  Figures~\ref{fig:lanes}c) and d) focus on how the ant distribution changes between each zone.

 \begin{figure}
 \centerline{\includegraphics[height=1.5in]{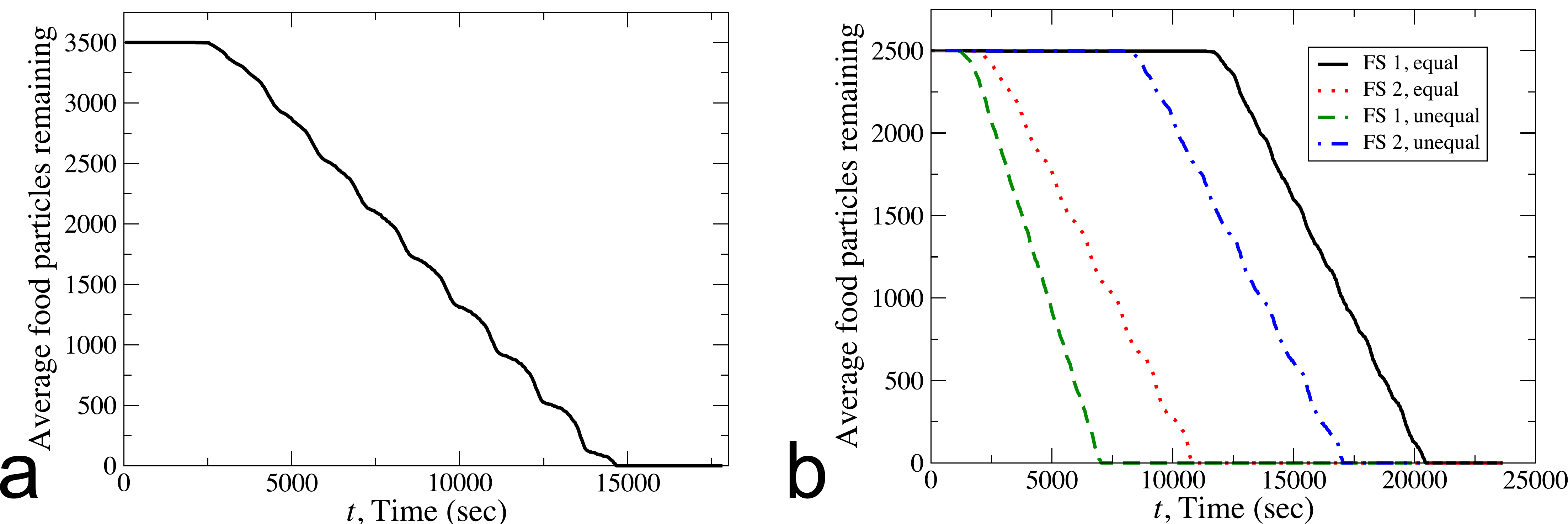}}
 \caption{Removal of food over the course of time for a) one food source or b) two food sources at equal and unequal distances from the nest.  Food depleted in 4-6 hours consistent with duration of raids from the experimental observations in \cite{Sch71,Sch40}.}\label{fig:food}
 \end{figure} 
 
 How can ants form bi-directional traffic lanes?  The model suggest it is the result of the excluded volume constraint and the break in symmetry between the social cues for the foragers and returners.  In addition, the lane size introduces an effective length scale, which is dictated by the particle size manifested in the truncated repulsive potential $U$ defined in \eqref{eqn:truncated}.

 This global traffic behavior is consistent with the previous theory and/or experiment in \cite{Cou03,Dus09,Fra85}, where ants self-organize into lanes for optimal transport of food back to the nest.  In addition, such traffic dynamics have also been recently observed in bacteria \cite{Ari13}.   Unlike the model presented in \cite{Cou03}, we do not impose a directional preference for half the ants, which may artificially contribute to the formation of the bi-modal distribution in that work.  Also, in \cite{Cou03} a turning parameter is used where outbound foragers have a higher avoidance rate, which essentially forces them to the outside of the trail.  Instead, in this work, the excluded volume forces alone arising from first principles naturally sort the ants.  Both models agree on the conclusion that lanes form due to the asymmetry  in interactions between foragers and returners.
 
 The second natural question posed concerns the formation of three lanes as opposed to two.  One explanation deduced from experimental observation in \cite{Cou03} is that a two lane flow would introduce a left-right asymmetry in the trail pattern not naturally present and thus limiting its efficiency.  There may be another explanation.  Our model suggests that the desire to return directly home with food outweighs the exponential decay of the chemical gradient away from the food source.  In nature, when ant is encumbered with food it wants to return to the colony as quickly as possible (verified experimentally in \cite{MuhWeh88,Weh03}).  When a foraging ant encounters a returning ant along a trail it is easier for the ant carrying nothing to move out of the way.  This can also be seen in Figure~\ref{fig:lanes} by noting that the trail width is approximately the size of one ant.  Alternatively, the first lane to appear and form is the central one for returners, which forms naturally in the middle to minimize the path back to the nest.  Once foragers detect the pheromone the returning lane has already formed and they have no choice but to step away from the middle to avoid collisions until they get very close to the food source.  When an ant returns to the nest with food and becomes a forager it has no bias to which side of the pheromone trail it will traverse.  This leads to the bi-modal distribution of foraging ants along side the main trail.  Similarly, when a returning ant leaves the food source it can be on either side of the trail explaining Figure~\ref{fig:lanes}d).

    \begin{figure}
 \centerline{\includegraphics[height=2.25in]{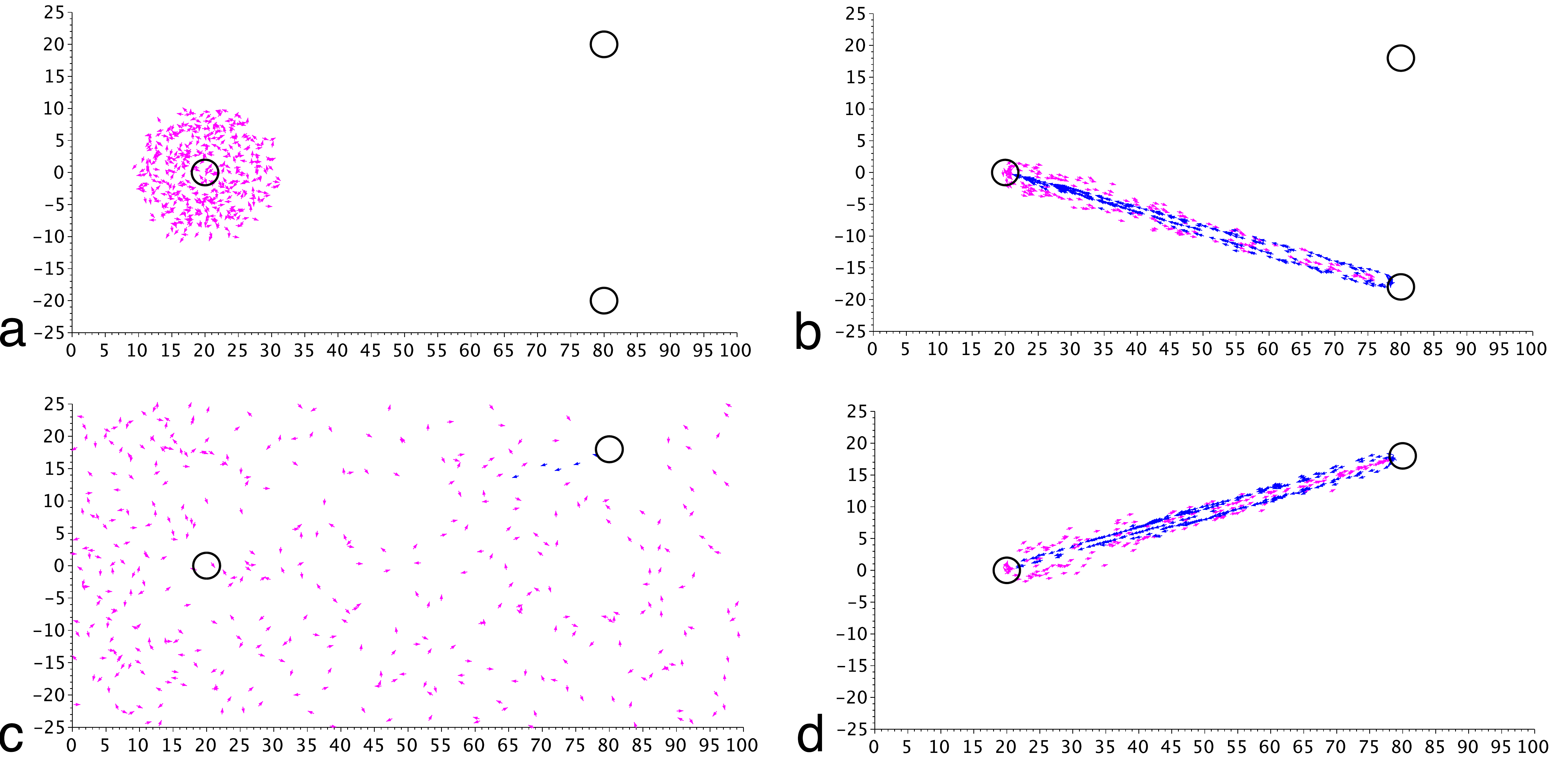}}
 \caption{Sample ant raiding simulations with foragers (purple) and returners (blue) where $N = 400$ and the food sources are equi-distant to the nest. Each arrow represents an individual ant's orientation ${\boldsymbol \omega}_i$.  The black circles denote the nest, ${\bf x}_c = (20,0)$, and food sources, ${\bf x}_{f,1} = (80,20)$ and ${\bf x}_{f,2} = (80,-20)$.  a) Initially ants are placed near the nest in non-overlapping positions with random orientation. b) The first food source is found and a trail develops similar to the case of only one food source.  c) Once that food source is depleted random foraging commences once again until the other food source is found. d) A trail forms at the second food source. See Online Resource 3.}\label{fig:trail2eq}
\end{figure} 
  
 \subsection{Trail disappearance} 
 
 Once the food source is exhausted the trail ceases to exist because the foraging ants no longer are attracted to it.  This behavior is captured by imposing a count on the quantity of food items (e.g., 2000-3500).  Once the food source is depleted no foragers can become returners and eventually the whole colony is composed of foragers looking for their next cache of resources.  For some insight into how the food is efficiently broken down and returned to the nest see Figure~\ref{fig:food}a) for the quantity of food particles as a function of time. Since this function has essentially a constant decreasing slope after the trail forms, one could argue the system has reached the maximally efficient state and remains there until the food is gone.  This provides further evidence for lane formation.  If lanes did not form one would expect regions of little decrease in Figure~\ref{fig:food}a) representing congestion along the trail. 

 After the trail disappears, $t > t_{dep}$, the simulations show that the ant distribution around the trail center for foragers becomes uniform and the lanes cease to exist (see Figure~\ref{fig:trail}d)).  Also, after the food has been depleted and the disordered state commences, one may notice local areas of milling behavior similar to \cite{Ber06}.  However, it is not well pronounced due to the presence of the random walk term in the dynamic equations.  
 Now we wish to extend our study to make predictions about the behavior with multiple food sources.

 \begin{figure}
 \centerline{\includegraphics[height=2.25in]{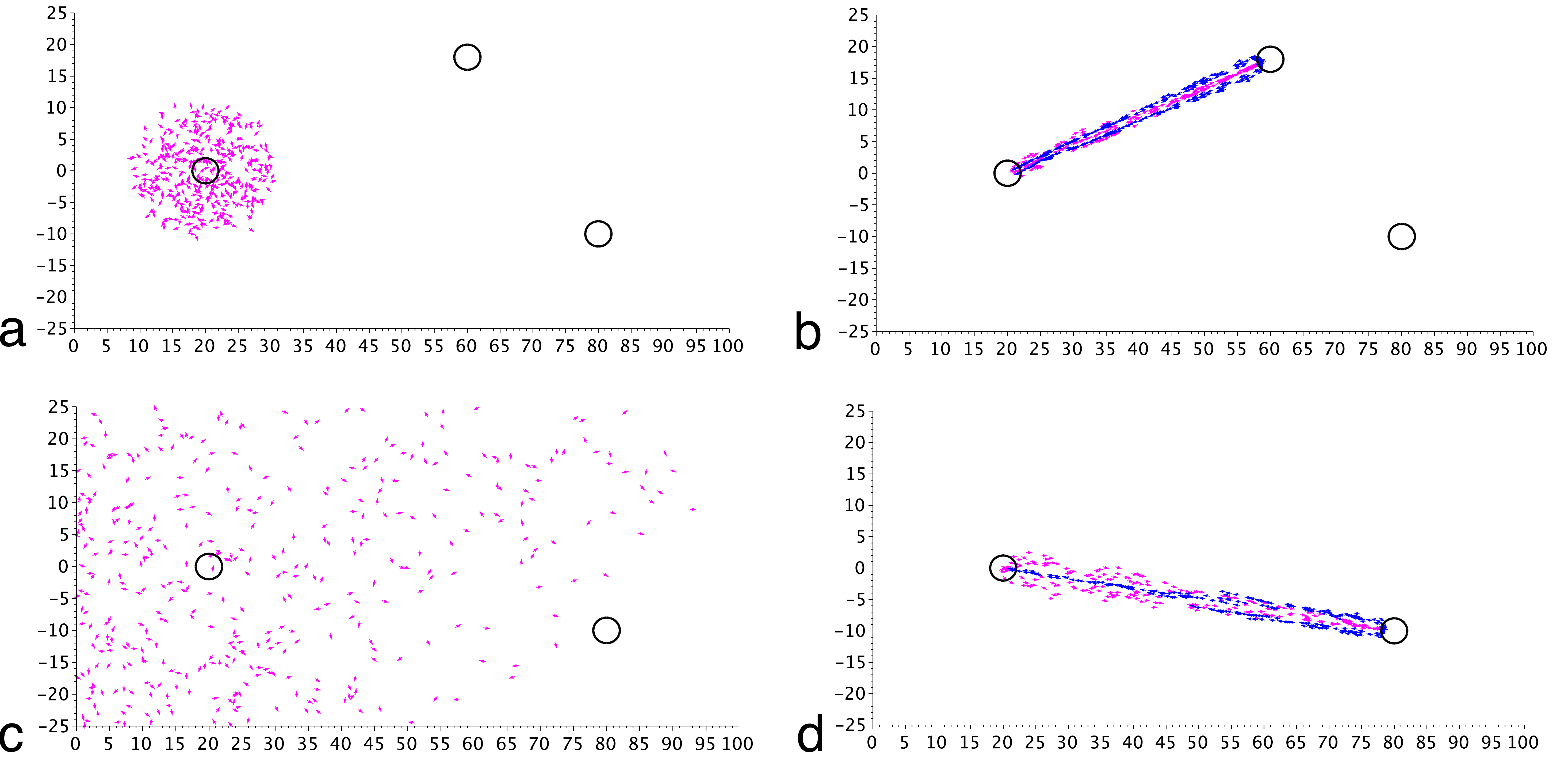}}
 \caption{Sample ant raiding simulations with foragers (purple) and returners (blue) where $N = 400$ and the food sources are equi-distant to the nest. Each arrow represents an individual ant's orientation ${\boldsymbol \omega}_i$.  The black circles denote the nest, ${\bf x}_c = (20,0)$, and food sources, ${\bf x}_{f,1} = (60,20)$ and ${\bf x}_{f,2} = (80,-10)$.  a) Initially ants are placed near the nest in non-overlapping positions with random orientation. b) The first food source is found and a trail develops similar to the case of only one food source.  c) Once that food source is depleted random foraging commences once again until the other food source is found. d) A trail forms at the second food source. See Online Resource 4.}\label{fig:trail2uneq}
\end{figure}

 \subsection{Multiple food sources}\label{sec:two_food}
 
 In this section, the transition to the collective state and local lane formation in the presence of multiple food sources is investigated.  Two main cases should be considered; namely, (i) two equidistant  and (ii) two non-equidistant food sources. In principle different foragers can find each food source near the same time.  Each will begin to deposit pheromone and return to the nest.  Naturally, foragers begin to detect whichever pheromone is closer to their current location and follow the trail to that food source leading to the formation of two distinct trails.  If the two food sources are at an equal distance from the nest one would expect the emergence of two near equivalent trails forming through the course of the raid.  In contrast, if one food source is significantly closer, one would expect most foragers to detect that pheromone sooner and the vast majority would complete the raid on the first food source before moving to the second. Multiple foraging locations as well as the study of a trail network have also been considered in \cite{Amo14,Sum03}.

 Both cases can be understood by analyzing the PDE for the pheromone concentration \eqref{eqn:pher-new}.   Since this equal is linear, multiple food sources can easily be considered by changing the righthand side to 
 \begin{equation*}
 \sum_{j=1}^{M_1} qe^{\|{\bf x}_j(t)-{\bf x}_{f,1}\|^2}\delta({\bf x}-{\bf x}_j(t)) + \sum_{p=1}^{M_2} qe^{\|{\bf x}_p(t)-{\bf x}_{f,2}\|^2}\delta({\bf x}-{\bf x}_p(t)).
 \end{equation*}
   If one food source is visited more frequently, then more terms in \eqref{eqn:fund2} will direct ants toward that food site.  We use simulations to study the distinct behavior among the two cases: (i) equidistant food sources (e.g., see Figure~\ref{fig:trail2eq} and Online Resource 3) and (ii) food sources at different distances (e.g., see Figure~\ref{fig:trail2uneq} and Online Resource 4).

\begin{figure}
 \centerline{\includegraphics[height=1.75in]{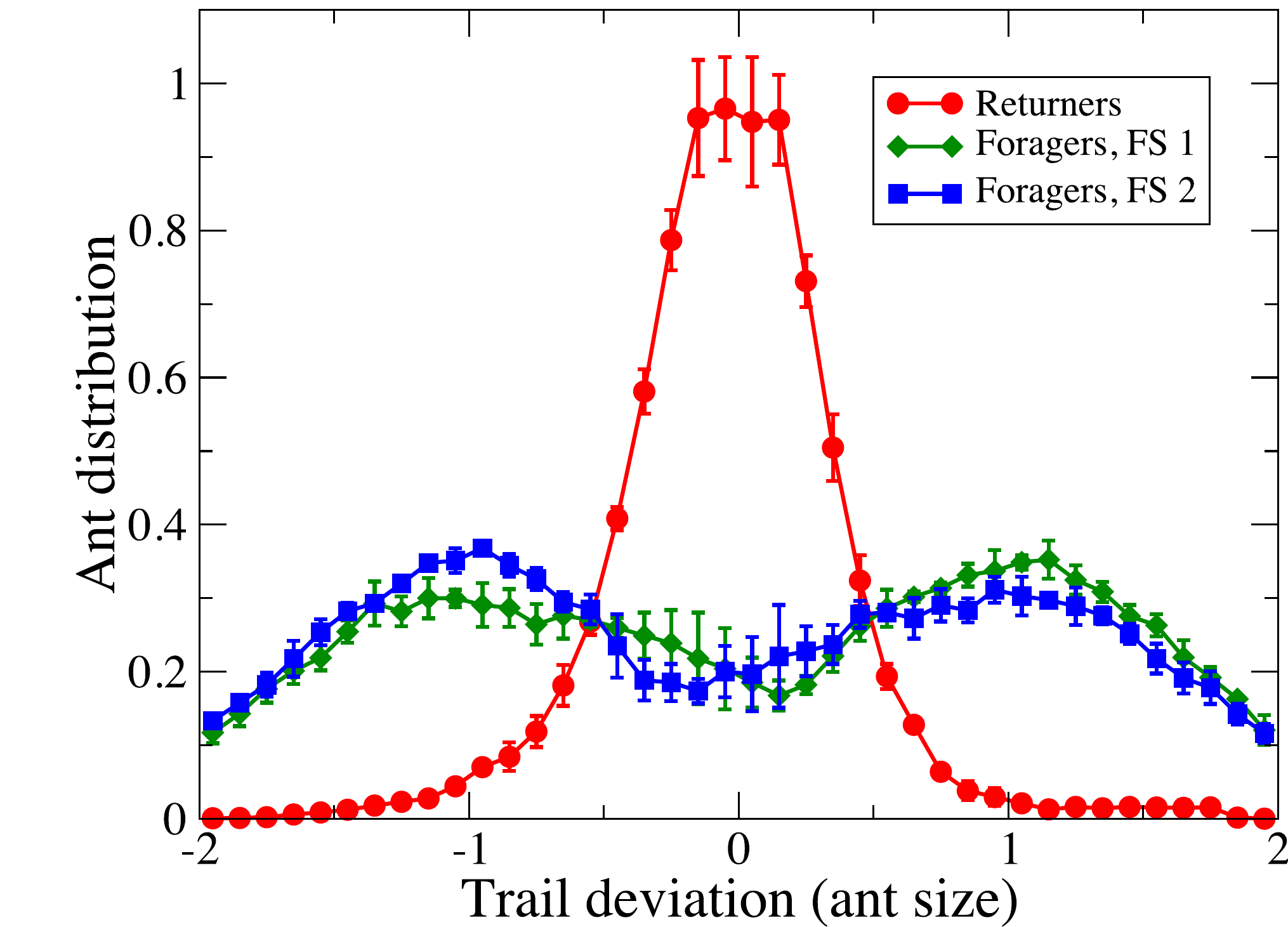}}
 \caption{Formation of lanes along the pheromone trail for foragers (green, blue) and returners (red).  Distances normalized by ant size $\ell=x_0$ (characteristic length).  Bi-modal peaks occur at a distance between $.5\ell-1.5\ell$ from the center indicating two outside lanes of foragers with returners in the middle.  Error bars represent one standard deviation.}\label{fig:lanes2}
 \end{figure}

In the former case, the foraging ants are equally probable to find either food source while completing the random walk.  We observe in simulations that both sites are visited initially, but the site that has more visitors eventually lures all the foraging ants due to the larger pheromone concentration.  The typical three lane local dynamics of foragers and returners can be observed on each trail (see Figure~\ref{fig:lanes2}).  In the case of food sources at unequal distances, the foraging ants find the closer food source first as they sweep across the computational domain.  Once a food source is found in either case almost all foragers are attracted to this site, which exhibits behavior similar to the one food source case (see Figures~\ref{fig:trail2eq},\ref{fig:trail2uneq}).  There is a period of random foraging again until the second food source is found.

 \begin{figure}
 \centerline{\includegraphics[height=1.65in]{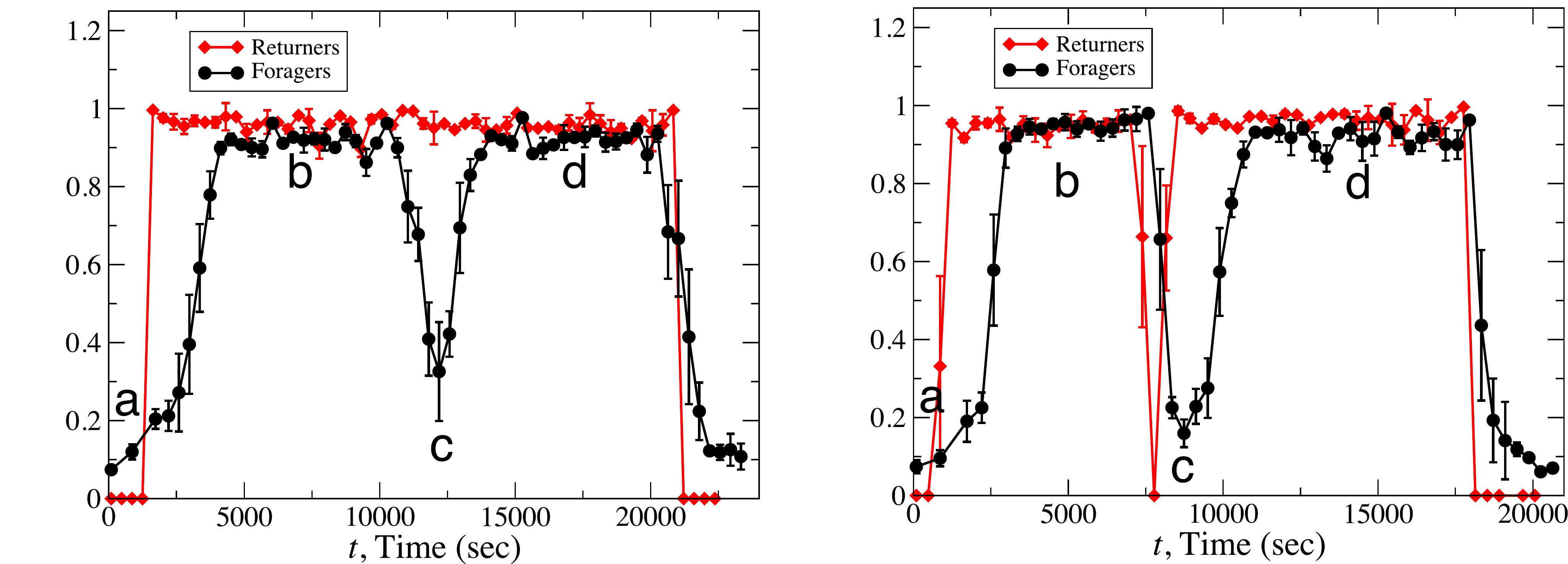}}
 \caption{Order parameter $F = \frac{1}{N}\left|\sum_{i=1}^N {\boldsymbol \omega}_i\right|$ versus time showing a transition from the individual to the collective state.  Left: Two equally spaced food sources.  Letters correspond to Figure~\ref{fig:trail2eq}a)-d). Right: Two no-equidistant food sources.  Letters correspond to Figure~\ref{fig:trail2uneq}a)-d). Error bars represent one standard deviation. Observe in the case of equally spaced food sources the raids last roughly the same amount of time indiciating a maximally efficient state has been reached and raid time only depends on the trail length.}\label{fig:trans2}
\end{figure}
 
 Next,  the model can be used to study the effect of two food sources on the transition to and duration of the collective state.  The main question is whether the system will form two coexisting collective states or one collective raid at the first food source and then another at the second.  We use the order parameter \eqref{eqn:op} for each class of ant to study the current state of the system.  Figure~\ref{fig:trans2}, shows that there is still a clear transition to the collective state in both cases; however there are a few subtle differences than in the case of one food source.  
 
 In both the cases of equidistant and non-equidistant food sources, a collective state is reached essentially in the same amount of time as the single food source case (see Figure~\ref{fig:trans2}).  This is due to the fact that once foragers find a food source they immediately lay pheromone attracting all other foragers nearby.  The result is single food source behavior until depletion where the foragers carry out a random walk again.  This can be seen explicitly in Figure~\ref{fig:trans2} during the time period that the foragers leave the collective phase.  This occurs because there was only a trace amount of chemical, if any, deposited at the second food source.  While the first food source was being raided this amount dissipated exponentially fast since foragers no longer visited.

Finally, we investigate the effect of multiple food sources on the lane formation.  Figure~\ref{fig:lanes2} show that the model captures the local traffic dynamics along each trail with the formation of the three lanes, two outside lanes for foragers and one internal lane for ants returning with resources.  This is consistent with the results observed in the case of one food source in Figure~\ref{fig:lanes}b) and is independent of the locations of the food sources.  
 
\section{Kinetic Phase Transition}~\label{sec:pt}

The model can also be used to study the behavior of the system near the transition to collective motion.  While we have loosely called this a ``phase transition" throughout this work, we must distinguish the definition used here from the classic one from thermodynamics.   The kinetic phase transition occurs when the order parameter exhibits behavior similar to that of a continuous phase transition in an equilibrium system \cite{Vic99,Vic95}.   The system under consideration here is far from equilibrium, yet still is capable of demonstrating phase transition type behavior.    For a rigorous treatment of phase transitions in systems of self-propelled particles consult \cite{Deg15}.

  \begin{figure}
 \centerline{\includegraphics[height=5.00in]{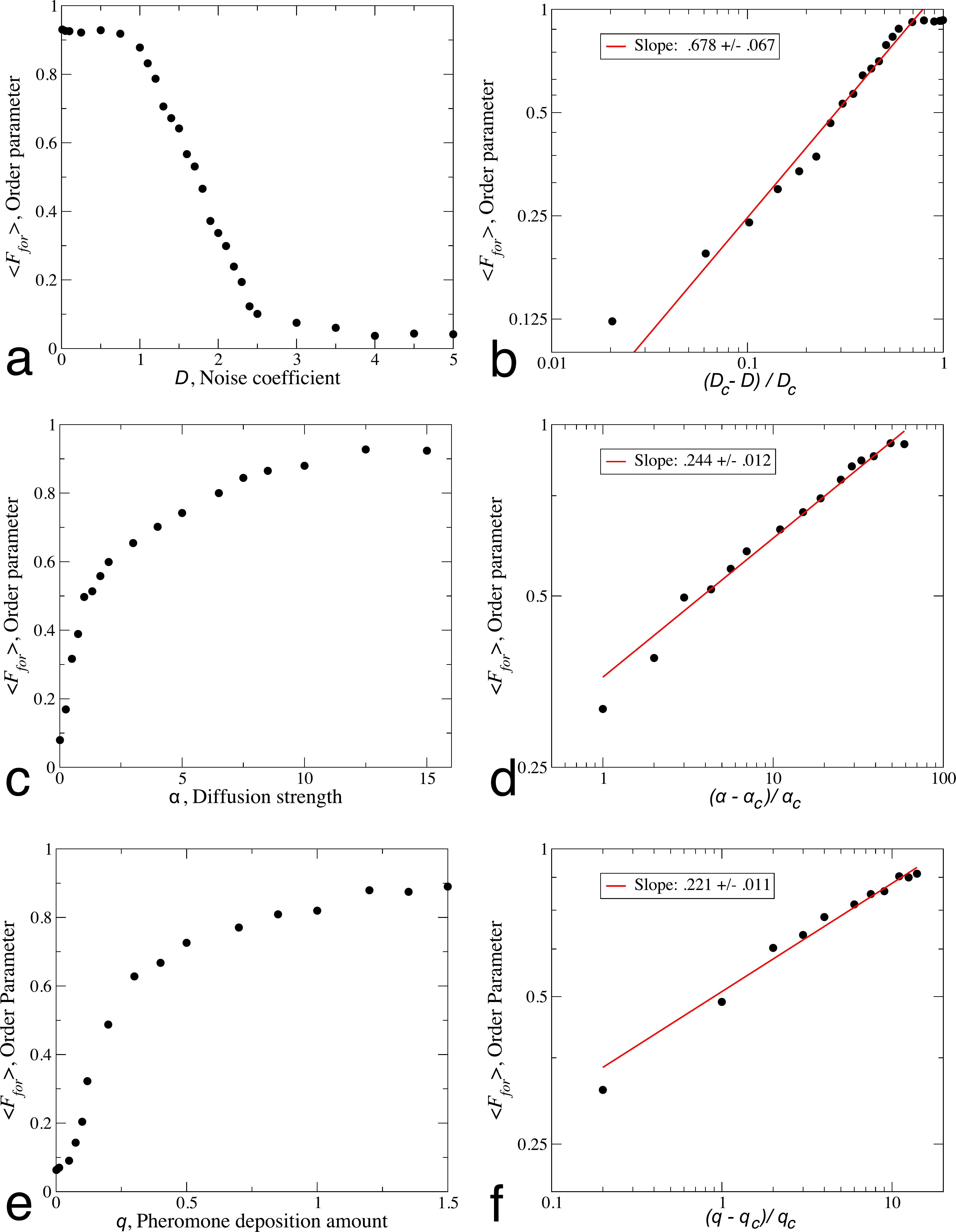}}
 \caption{The avergage value of the order parameter over the course of the raid, $\langle F_{for}\rangle$, exhibits a continuous kinetic phase transition as a function of the system parameters for a) the noise strength $D$, c) the rate of diffusion $\alpha$, and e) the phermone deposition amount $q$.  The critical exponents are extracted from the corresponding log-log plots of $\langle F_{for}\rangle$.}\label{fig:pt}
 \end{figure}

 In this work, we will follow the approach presented in \cite{Vic99,Vic95} where the kinetic phase transition from no transport (e.g., order parameter $F = 0$) to finite net transport as a function of system parameters was first studied.  If one can find a critical exponent $\zeta$ such that $F \sim (\eta_c - \eta)^{\zeta}$, then the system is said to posses a {\it continuous kinetic phase transition}. Here $\eta$ is the system parameter under investigation (e.g., strength of noise or density as in \cite{Vic99,Vic95}) and $\eta_c$ is the critical value of this parameter near the transition.   The behavior near the transition is referred to as {\it self-organized critical behavior} because it spontaneously arises in dissipative systems due to the dynamics of interacting units \cite{Jen98,Win15}.  By deriving a power law the system is considered to be ``scale-free" and is therefore universal \cite{New06}.  This field combines self-organization and critical behavior to provide greater understanding of the complexity of a given system of interacting units \cite{Jen98}.
 
 Specifically, in this work, we study the average value of the order parameter for foragers, $\langle F_{for} \rangle$, during the raiding time period  as the noise strength in the random walk $D$, the diffusion coefficient $\alpha$, and the maximum amount of pheromone deposited $q$ vary.  Figure~\ref{fig:pt}a),c),e) show $\langle F_{for} \rangle$ as a function of these system parameters.  In each variable we estimate the point where self-organization begins (e.g., $D_c = 2.45$, $\alpha_c = .4$, and $q = .04$).  Using a log-log plot the data is fit to a power law where the slope gives the critical exponent (see Figure~\ref{fig:pt}b),d),f))
 \begin{equation*}
 F_{for} \sim (D_c - D)^{\beta_1}, \quad F_{for} \sim (\alpha - \alpha_c)^{\delta_1}, \quad F_{for} \sim (q - q_c)^{\delta_2}.
 \end{equation*} 
 From the data we find critical exponents $\beta_1 \approx 2/3$, $\delta_1 \approx 1/4$, and $\delta_2 \approx 1/4$.   These critical exponents can also be observed in thermodynamic systems found in nature.  For example, the magnetization in a spin system is proportional to the applied field, $m \propto h^{\delta}$ where $\delta \approx .25$ \cite{Ma76}.  In addition, a critical exponent near $2/3$ is observed when considering the density of a superfluid ($He$) versus temperature, $\rho_{s} \propto (T_c-T)^\beta$ for $\beta \approx .667$ \cite{Sta71}.  For comparison $\beta = 1/2$ is the mean-field critical exponent for the order parameter as a function of temperature in an Ising-like model from classical thermodynamics \cite{Sta71}.   The fact that the derived critical exponents match some classic thermodynamics systems speaks to the universality of the model and explains why this behavior is referred to as a phase transition.  These plots are sensitive to the choice of critical value and, therefore, these exponents can only be taken as approximations.  However, the exact determination of the critical exponents is beyond the scope of the present work and Figure~\ref{fig:pt} was presented only illustrate that the system does exhibit a continuous kinetic phase transition.

 \section{Discussion}~\label{sec:disc}
  
\subsection{Limitations and future work}

While the model uses microscopic interactions to accurately capture the macroscopic state, it still has some limitations. One limitation is the numerical approximation of the time integral in \eqref{eqn:fund}.  One could overcome this with a pure diffusion model as outlined in Appendix~\ref{app:diff}, but this is only sufficient on short trails.   In \cite{Deg13}, the deposition of pheromones along a trail is incorporated via a kinetic model and the simulations are carried out by solving this PDE directly, but the entire ant raiding cycle (e.g., foraging and returning to the nest) has not yet been studied.    


Another simplification used in this work is a homogeneous environment where essentially there are no obstacles or variance in elevation.  Some progress has been made toward this in \cite{Col92}, but much more is still needed for full understanding.  It would be interesting to include environmental effects and study how the onset of collective dynamics and lane formation would change.  One could add a spatial dependence to $\xi = \xi({\bf x})$ in the self-propulsion term to model the frictional component from Rayleigh's Law as suggested in \cite{CarForTosVec10}.  If the environment is rough and movement is hindered ($\xi$ is small) or if the environment is flat and homogeneous with little friction, then $\xi$ is closer to the isolated translational speed of an ant.   Since the model accounts for interactions between ants in one colony, it is natural to ask what would happen if this model was used for  ant communities competing for food resources \cite{Mar15,Pow04} or if the effect of predators was investigated \cite{Kas03,Cou05}.   

The clear next step is to derive the corresponding kinetic theory for the coupled system \eqref{eqn:pher}-\eqref{eqn:ibm-r} for comparison with the various current continuum PDE models for the ant density.  There the limit is taken as the number of particles and volume go to infinity, but the concentration $N/|V_L|$ is held fixed.   Then it may be possible to establish existence and uniqueness results through mathematical analysis using techniques from ODE and PDE theory (e.g., similar to what has been done for bacterial suspensions \cite{Rya13a} and locusts \cite{Erb12,Esc10}).

Finally, the implementation of the elliptical truncated Gay-Berne potential in place of the isotropic truncated Lennard-Jones potential \eqref{eqn:truncated} would allow one to study the effect of ant shape as discussed in Remark 1.  Since ants are elongated with aspect ratios between 3-6, the shape may play some role in the near-field collisional interactions affecting how the lanes form and their intrinsic size.

 \subsection{Conclusions}

This work introduced a new coupled PDE/ODE model for pheromone concentration and ant dynamics.  Through analysis of the model,  one can study the physical parameters governing the transition to and from a collective state that occurs during the course of a raid.  While experiments have limitations including observation time and lack of control over the individuals, the mathematical model introduced allows for a deep study of the dynamics of the raid for long periods of time.  The model has been verified to match the qualitative behavior observed in the experimental results of \cite{Cou03,Sch71,Sch40}.  

The main result of this work is the introduction of a new model capable of capturing the emergence and dissipation of an ordered state as well as the self-organization of individuals into traffic lanes for efficient transport of resources.  The simulations of the model show indeed a sharp transition from individual to collective behavior in both foragers and returners with an explicit time delay accounting for the reaction of the foragers to the chemical gradient.   Also, the model reveals that the lanes form due to the presence of an excluded volume constraint and collisions. The case of multiple food sources was investigated revealing distinct behavior depending on the locations of the food sources.  Also, the critical exponents found herein describe the behavior of the system as a function of system parameters.

Through analysis of the model, we acquire further knowledge of social insect behavior.  Even in the absence of direct communication, the model shows that ants can still self-organize into efficient transport pathways.  This is the result of a complex network of chemical signaling through pheromone detection and deposition as well as local near-field collision avoidance.  While ants are one example of social insects, the nonverbal cues are present in other species.  In absence of verbal communication humans at a crosswalk unconsciously form lanes for efficient travel.  This can be explained using insight from the analysis of the model for ants in that an individual takes up a certain amount of space and to avoid a path being inhibited  individuals of like orientation naturally follow one another.  As in ants, the global patterns are not known at the local level, yet still emerge in time.   The main difference between humans and ants is that typically humans behave  in a ways that are best for the individual while ants only exist for the good of the colony \cite{Cou03}.

At present this work only considers a small number of ants to verify the model.  This allows for figures and simulations where the particles can be distinguished to illustrate individual behavior at the microscopic level.  Even with this restriction in mind, this simple model is still able to capture the transition to the collective state and lane formation.  Most other works focus on one aspect of the raiding cycle such as laying a chemical trail. However, our model, like the recent PDE model in \cite{Amo14},  allows for simulation of the entire ant raiding cycle from random foraging, to the identification of a food source, and food depletion.   Once the food is gone the model naturally accounts for the degradation of the trail with the disappearance of the pheromone chemical gradient exponentially fast and the transformation of all ants to foragers.

Overall, the model introduced in this work provides novel insight into the raiding behavior of ants while laying the foundation for investigating future questions such as elevation effects, competing colonies, and predators.  The simple nature of the model only keeps the necessary biological parameters needed to reach the ordered state reducing the study of a complex phenomena to a system of interacting points governed by a  balance of forces.  This work highlights the interplay between two communities of ants within the same colony in order to achieve an efficient state of resource transport fundamental to daily life.
 
 \thanks{{\bf Acknowledgments} Thank you to Paulo Amorim, Gil Ariel, and Magali Tournus for useful discussions.  The author gratefully acknowledges support from National Science Foundation Grant DMS-1212046 and advice from X. Zheng (KSU) and P. Palffy-Muhoray (KSU).}
 
\bibliographystyle{spbasic}

\begin{thebibliography}{0}
\providecommand{\natexlab}[1]{#1}
\providecommand{\url}[1]{{#1}}
\providecommand{\urlprefix}{URL }
\expandafter\ifx\csname urlstyle\endcsname\relax
  \providecommand{\doi}[1]{DOI~\discretionary{}{}{}#1}\else
  \providecommand{\doi}{DOI~\discretionary{}{}{}\begingroup
  \urlstyle{rm}\Url}\fi
\providecommand{\eprint}[2][]{\url{#2}}

\end{thebibliography}


\begin{thebibliography}{2}

\bibitem{Amo14} Amorim P. (2014) Modeling ant foraging: {A} chemotaxis approach with pheromones and trail formation. arXiv  \url{http://arxiv.org/pdf/1409.3808.pdf}

\bibitem{Ara13} Aranson I.~S., (2013) Collective behavior in out-of-equilibrium colloidal suspensions. C. R. Physique 14:518-527

\bibitem{Ari14} Ariel G., Ophir Y., Levi S., Ben-Jacob E., Ayali A., (2014) Individual pause-and-go motion is instrumental to the formation and maintenance of swarms of marching locust nymphs.  PLOS One 9(7):e101636

\bibitem{Ari13} Ariel G., Shklarsh A., Kalisman O., Ingham C., Ben-Jacob E., (2013) From organized internal traffic to collective navigation of bacterial swarms.  New Journal of Physics 15:125019 


\bibitem{Bat63} Bates H.~W., (1863) The naturalist on the river {A}mazons. London: Murray 2: 350-366  

\bibitem{Bec92} Beckers R., Deneubourg J.~L., Goss S. (1992) Trail laying behavior during food recruitment in the ant Lasius niger ({L}.).  Springer Insectes soc. 39(1):59-72


\bibitem{Deg13} Boissard E., Degond P., Motsch S., (2013) Trail formation based on directed pheromone deposition. Journal of Mathematical Biology 66(6):1267-1301

\bibitem{Bue14} Buehlmann C., Graham P., Hansson B.~S., Knaden M., (2014) Desert ants locate food by combining high sensitivity to food odors with extensive crosswind runs.  Current Biology 24: 960-964

\bibitem{Bur12} Burger M., Haskovec J., Wolfram M.-T., (2012) Individual-based and mean-field modelling of direct aggregation. Physica D - Nonlinear Phenomena 260

\bibitem{Cam97} Bonabeau E., Theraulaz G., Deneubourg J.-L., Aron S., Camazine S., (1997) Self-organization in social insects. TREE 12(5): 188-193

\bibitem{Cal92} Calenbuhr V., Deneubourg J.-L., (1992) A model for osmotropotactic orientation (I).  J. Theor. Biol. 158: 343-349


\bibitem{Car09} Carillo  J.~A., D'Orsogna M.~R., Panferov V., (2009) Double milling in self-propelled swarms from kinetic theory. Kinetic and Related Models 2(2):363-378 

\bibitem{CarForTosVec10} Carillo J.~A., Fornasier M., Toscani G., Vecil F., (2010) Particle, kinetic, and hydrodynamic models of swarming. Mathematical Modeling of Collective Behavior in Socio-Economic and Life Sciences. Birkh\"{a}user Boston 297-336

\bibitem{Col92} Colorni A., Dorigo M., Maniezzo V., (1992) Distributed optimization by ant colonies.  Proc. of ECAL91 Paris, France 134-142

\bibitem{Cou03} Couzin I.~D., Franks N.~R., (2003) Self-organized lane formation and optimized traffic flow in army ants. Proc. R. Soc. Lond. B 270:139-146

\bibitem{Vic99} Czir\'{o}k A., Barab\'{a}si A.-L., Vicsek T., (1999) Collective motion of self-propelled particles: {K}inetic phase transition in one dimension.  Phys. Rev. Letters 82(1):209-212
 
\bibitem{Ber06} D'Orsogna M.~R., Chuang Y.~L., Bertozzi A.~L., L.~S. Chayes, (2006) Self-propelled particles with soft-core interactions: {P}atterns, stability, and collapse.  Phys. Rev. Letters 96:104302

\bibitem{Deg15} Degond P., Frouvelle A., Liu J.-G., (2015) Phase transitions, hysteresis, and hyperbolicity for self-organized alignment. Arch. Rational Mech. Anal. 216:63-115

\bibitem{Dus09} Dussutour A., Beshers S., Deneubourg J.-L., Fourcassie V, (2009) Priority rules govern the organization of traffic on foraging trails under crowding conditions in the leaf-cutting ant {\it Atta colombica}.  J. Exper. Biol. 212:499-505

\bibitem{Erb12} Erban R., Haskovec J., (2012) From individual to collective behaviour of coupled velocity jump processes: {A} locust example.  Kinetic and Related Models 5(4)

\bibitem{Esc10} Escudero C., Yates C., Buhl J., Couzin I., Erban R., Kevrekidis I., Maini P., (2010) Ergodic directional switching in mobile insect groups. Physical Review E 82(1):011926


\bibitem{Fra85} Franks N.~R. (1985) Reproduction, foraging efficiency and worker polymorphism in army ants. Experimental behavioral ecology Stuttgart: G. Fisher 97-107

\bibitem{Gar13} Garnier S., Combe M., Jost C., Theraulaz G., (2013) Do ants need to estimate the geometrical properties of trail bifurcations to find an efficient route? {A} swarm robotics test bed. PLoS Computational Biology 8(3):e1002903

\bibitem{Gar09} Garnier S., Gu\'{e}r\'{e}cheau A., Combe M., Fourcassi\'{e} V., Theraulaz G., (2009) Path selection and foraging efficiency in {A}rgentine ant transport networks Behav Ecol Sociobiol (2009) 63:1167-1179

\bibitem{Got95} Gotwald W.~H. (1995) Army ants: {T}he biology of social predation. Cornell University Press Ithaca, NY

\bibitem{Hol90} H\"{o}lldobler B., Wilson E.~O. (1990) The ants. The Belknap Press of Harvard University Press, Cambridge, Mass

\bibitem{Jen98} Jensen H.~J., (1998) Self-organized criticality. Cambridge University Press, New York, NY 

\bibitem{Joh06} Johnson K., Rossi L.~F. (2006) A mathematical and experimental study of ant foraging trail dynamics. Journal of Theoretical Biology 241:360-369

\bibitem{Kas03} Kaspari M., O'Donnell S. (2003) High rates of army ant raids in the {N}eotropics and implications for ant colony and community structure. Evolutionary Ecology Research 5:933-939

\bibitem{Kel70} Keller E., Segel L. (1970) Initiation of slide mold aggregation viewed as an instability. J. Theoretical Biology 241:360-369

\bibitem{Kel71} Keller E., Segel L. (1971) Model for chemotaxis.  J. Theoretical Biology 26:399-415

\bibitem{She12} Lushi E., Goldstein R.~E., Shelley M.~J. (2012) Collective chemotactic dynamics in the presence of self-generated fluid flows. Physical Review E 86: 040902(R)

\bibitem{Ma76} Ma S.-K., (1976) Modern theory of critical phenomena.  Benjamin, Reading

\bibitem{Mar15} Martelloni G., Santarlasci A., Bagnoli F., Santini G. (2015) Modeling ant battles by means of a diffusion-limited {G}illespie algorithm. arXiv preprint. 

\bibitem{MuhWeh88} M\"{u}ller M., Wehner R. (1988) Path integration in desert ants, {\it Cataglyphis fortis}. Proc. Natl. Acad. Sci. USA 85: 5287-5290

\bibitem{Nar13} Narendra A., Gourmaud S., Zeil J., (2013) Mapping the navigational knowledge of individually foraging ants, {\it Myrmecia croslandi}. Proc. R. Soc. B 280:20130683

\bibitem{New06} Newman, M.~E.~J., (2005) Power laws, {P}areto distributions and {Z}ipf's law. Contemporary physics 46(5):323-351

\bibitem{Per12} Perna A., Granovskiy B., Garnier S., Nicolis S.~C., Lab\'{e}dan M., Theraulaz G., Fourcassi\'{e} V., Sumpter D.~J.~T., (2012) Individual rules for trail pattern formation in {A}rgentine ants ({L}inepithema humile).  PLoS Comput biol 8(7):e1002592

\bibitem{Pow04} Powell S., Clark E. (2004) Combat between large derived societies: {A} subterranean army ant established as a predator of mature leaf-cutting ant colonies.  Insect. Soc. 51:342-351

\bibitem{Qi12} Qi W., Xu Y., Yung K.-L., Chen Y. (2012) A modified Gay-Berne model for liquid crystal molecular dynamics simulation. Polymer 53: 634-639


\bibitem {Rya11} Ryan S.~D., Haines B.~M., Berlyand L., Ziebert F., Aranson I.~S. (2011) Viscosity of bacterial suspensions: {H}ydrodynamic interactions and self-induced noise.  Physical Review E 83:050904(R) 

\bibitem{Rya13a} Ryan S.~D., Berlyand L., Haines B.~M., Karpeev D.~A. (2013) A kinetic model for semidilute suspensions.  SIAM Multiscale Modeling and Simulations 11(4):1176-1196

\bibitem{Rya13b} Ryan S.~D., Sokolov A., Berlyand L., Aranson I.~S. (2013) Correlation properties of collective motion in bacterial suspensions. New Journal of Physics 15:105021

\bibitem{She13} Saintillan D., Shelley M.~J., (2013) Active suspensions and their nonlinear models. C. R, Physique 14:497-517

\bibitem{Sch10} Schmickl T., Thenius R., Crailsheim K., (2010) Swarm-intelligent foraging in honeybees: benefits and costs of task-partitioning and environmental fluctuations. Neural Comput \& Applic 21:251-268

\bibitem{Sch71} Schneirla T.~C. (1971) Army ants: {A} study in social organization. Freeman San Francisco, CA

\bibitem{Sch40} Schneirla T.~C. (1940) Further studies of the army-ant behavior pattern. Mass organization in the swarm-raiders. Journal of Comparative Psychology 29(3):401

\bibitem{Sch97} Schweitzer F., Lao K., Family F. (1997) Active random walkers simulate trunk trail formation by ants. Biosystems 41:153-166

\bibitem{Ari11} Shklarsh A., Ariel G., Schneidman E., Ben-Jacob E., (2011) Smart swarms of bacteria-inspired agents with performance adaptable interactions.  PLOS Computational Biology 7(9):e1002177

\bibitem{Sol00} Sole R.~V., Bonabeau E., Delgado J., Fern\'{a}ndez P., Mar\'{i}n J., (2000) Pattern formation and optimization in army ant raids.  Artificial Life 6:219-226

\bibitem{Sta71} Stanley H.~E., (1971) Introduction to phase transitions and critical phenomena. Oxford University Press, Oxford

\bibitem{Sum12} Sumpter D.~J.~T., Mann R.~P., Perna A., (2012) The modelling cycle for collective animal behaviour.  Interface Focus 2:764-773

\bibitem{Sum03} Sumpter D.~J.~T., Beekman M., (2003) From nonlinearity to optimality: {P}heromone trail foraging by ants.  Animal Behaviour 66:273-280

\bibitem{Tsc06} Tschinkel W.~R. (2006) The fire ants. Harvard University Press

\bibitem{Vic95} Vicsek T., Czir\'{o}k A., Ben-Jacob E., Cohen I., Shochet O., (1995) Novel type of phase transition in a system of self-driven particles.  Physical Review Letters 75(6):1226

\bibitem{Vic12} Vicsek T., Zafeiris A., (2012) Collective Motion. Physics Reports 517:71-140

\bibitem{Vit06} Vittori K., Talbot G., Gautrais J., Fourcassi\'{e} V., Ara\'{u}jo A.~F.~R., Theraulaz G., (2006) Path efficiency of ant foraging trails in an artificial network. Journal of Theoretical Biology 239:507-515

\bibitem{Wat95a} Watmough J., Edelstein-Keshet L. (1995) A one-dimensional model of trail propagation by army ants. J. Math Biology 33:459-476

\bibitem{Wat95b} Watmough J., Edelstein-Keshet L. (1995) Modelling the formation of trail netwrosk by foraging ants. J. Theoretical Biology 176:357-371

\bibitem{Weh03} Wehner R. (2003) Desert ant navigation: how miniature brains solve complex tasks. J. Comp Physiol A 189:579-588

\bibitem{Wil62} Wilson E.~O. (1962) Chemical communication among workers of the fire ant {S}olenopsis saevissima (Fr. Smith) 1. The organization of mass-foraging.  Animal Behavior 10(1-2):134-138 

\bibitem{Win15} Winkler M., Falk J. Kinzel W. (2015) On the effect of the drive on self-organized criticality. arXiv:1410.5712v3

\bibitem{Cou05} Wrege P.~H., Wikelski M., Mandel J.~T., Rassweiler T., Couzin I.~D., (2005) Antbirds parasitize foraging army ants.  Ecology 86(3): 555-559

\bibitem{Cou05} Wrege P.~H., Wikelski M., Mandel J.~T., Rassweiler T., Couzin I.~D., (2005) Antbirds parasitize foraging army ants.  Ecology 86(3): 555-559

\bibitem{Xue13} Xue C., (2015) Macroscopic equations for bacterial chemotaxis: integration of detailed biochemistry of cell signaling. Journal of Mathematical Biology 70(1-2):1-44


\end{thebibliography}

\begin{appendix}

 \section{Non-dimensionalization}\label{app:nd}
 
In order to form a dimensionless problem for the purpose of numerical computations, we must now introduce characteristic scales.  A characteristic scale will be denoted by a subscript zero (e.g., $x_0$) and a non-dimensional quantity will be denoted with a hat.  For example 
\begin{equation*}
x = x_0\hat{x}, \quad t = t_0\hat{t}
\end{equation*}
where the characteristic size of an ant $x_0 = 1 \text{ cm}$ and one option for the characteristic time $t_0 = 101 s$ is based on the half-life of a food source taken from \cite{Amo14}.  The characteristic diffusion coefficient for pheromone is $\alpha_0 =x_0^2/t_0 = .01 \text{cm}^2/\text{s}$ and the characteristic concentration of pheromone deposited on a 2D surface is $c_0 = 1.1\times 10^{-4} \text{g cm}^{-2}$ (both match experimental values from \cite{Cal92,Cou03}).  

First, the homogeneous version of the PDE for pheromone concentration \eqref{eqn:pher} becomes
\begin{equation*}
\begin{cases}
\frac{c_0}{t_0}\partial_{\hat t} \hat{c} - \frac{\alpha_0 c_0\hat{\alpha}}{x_0^2}\Delta \hat{c} + \gamma c_0 \hat{c} = 0, & \quad \hat{\bf x} \in \mathbb{R}^2, \hat{t} \in (0, \infty)\\
c_0\hat{c}({\bf x}, 0) = c_0\hat{g}(\hat{\bf x}), & \quad \hat{\bf x} \in \mathbb{R}^2
\end{cases}.
\end{equation*}
By multiplying through the first equation by $t_0/c_0$ and the second by $1/c_0$ we find a non-dimensional equation for the concentration with non-dimensional parameters $\hat{\alpha},\hat{\gamma}$, and $\hat{q}$.  Once done, we replace the source term responsible for the exponential decay of the pheromone along the trail in dimensionless form
\begin{equation*}
\begin{cases}
\partial_{\hat t} \hat{c} - \hat{\alpha}\Delta \hat{c} + \hat{\gamma}\hat{c} = \sum_{j=1}^M\hat{q}e^{-\|\hat{\bf x}_j(t)-\hat{\bf x}_f\|^2}\delta(\hat{\bf x}-\hat{\bf x}_j(t)), & \quad \hat{\bf x} \in \mathbb{R}^2, \hat{t} \in (0, \infty)\\
\hat{c}({\bf x}, 0) = \hat{g}(\hat{\bf x}), & \quad \hat{\bf x} \in \mathbb{R}^2.
\end{cases}
\end{equation*}
where $\hat{\gamma} = \gamma t_0$ ($\gamma$ has units of 1/sec, $\gamma \approx 1/300s$ in \cite{Cou03}).  The maximal dimensionless concentration of pheromone deposited is $\hat{q} = q/c_0$.  Next, we proceed to the equations for the foraging ants without the white noise term
\begin{equation*}
\begin{cases}
\frac{x_0}{t_0}\dot{\hat{\bf x}}_i &= \frac{x_0}{t_0}\hat{\bf v}_i\\
\frac{x_0}{t_0^2}\dot{\hat{\bf v}}_i &=  \frac{x_0^3}{t_0^3}\nu\hat{\bf v}_i\left(\hat{\xi}^2-|\hat{\bf v}_i|^2\right) - \frac{1}{Nx_0}\sum_{j \neq i} \nabla_{\bf x} U(|\hat{\bf x}_i - \hat{\bf x}_j|) + dc_0\nabla_{\hat{\bf x}}\hat{c}(\hat{\bf x},\hat{t})
\end{cases}
\end{equation*}
By multiplying through the first equation by $t_0/x_0$ and the second by $t_0^2/x_0$ we find a non-dimensional equation and add in the dimensionless Gaussian white noise
\begin{equation*}
\begin{cases}
\dot{\hat{\bf x}}_i &= \hat{\bf v}_i\\
\dot{\hat{\bf v}}_i &=  \hat{\nu}\hat{\bf v}_i\left(\hat{\xi}^2-|\hat{\bf v}_i|^2\right) - \frac{1}{N}\sum_{j \neq i} \nabla_{\bf x} \hat{U}(|\hat{\bf x}_i - \hat{\bf x}_j|) + \hat{d}\nabla_{\hat{\bf x}}\hat{c}(\hat{\bf x},\hat{t})+\hat{D}\hat{W}_{\hat{t}}.
\end{cases}
\end{equation*}
where $\hat{U}$ is defined in \eqref{eqn:truncated} with dimensionless relative distance $\hat{r}$ and the dimensionless depth of the potential well $\hat{\ve} = \ve_0 t_0^2/x^2_0$. Also, $\hat{\nu} = \nu x_0/t_0$ and $\hat{d} = dc_0t_0^2/x_0^2$.  Similarly for returning ants we find
\begin{equation*}
\begin{cases}
\dot{\hat{\bf x}}_i &= \nu\hat{\boldsymbol \omega}_i\\
\dot{\hat{\bf v}}_i &= \hat{\nu}\hat{\bf v}_i\left(\hat{\xi}^2-|\hat{\bf v}_i|^2\right) - \frac{1}{N}\sum_{j \neq i} \nabla_{\bf x} \hat{U}(|\hat{\bf x}_i - \hat{\bf x}_j|) + \hat{\beta}\frac{\hat{\bf x}_i - \hat{\bf x}_c}{\hat{r}}
\end{cases}
\end{equation*}
where $\hat{\beta} = \beta t_0^2/x_0$ and $\hat{r} = |\hat{\bf x}_i-\hat{\bf x}_c|$.  Even though the model \eqref{eqn:ibm-f}-\eqref{eqn:ibm-r} was formulated with dimensional constants, from the dimensional analysis presented in this appendix we recover the necessary dimensions of each of the original quantities if desired.  Throughout this work the hats are dropped and all variables are understood as dimensionless.  

 \begin{figure}
 \centerline{\includegraphics[height=2.75in]{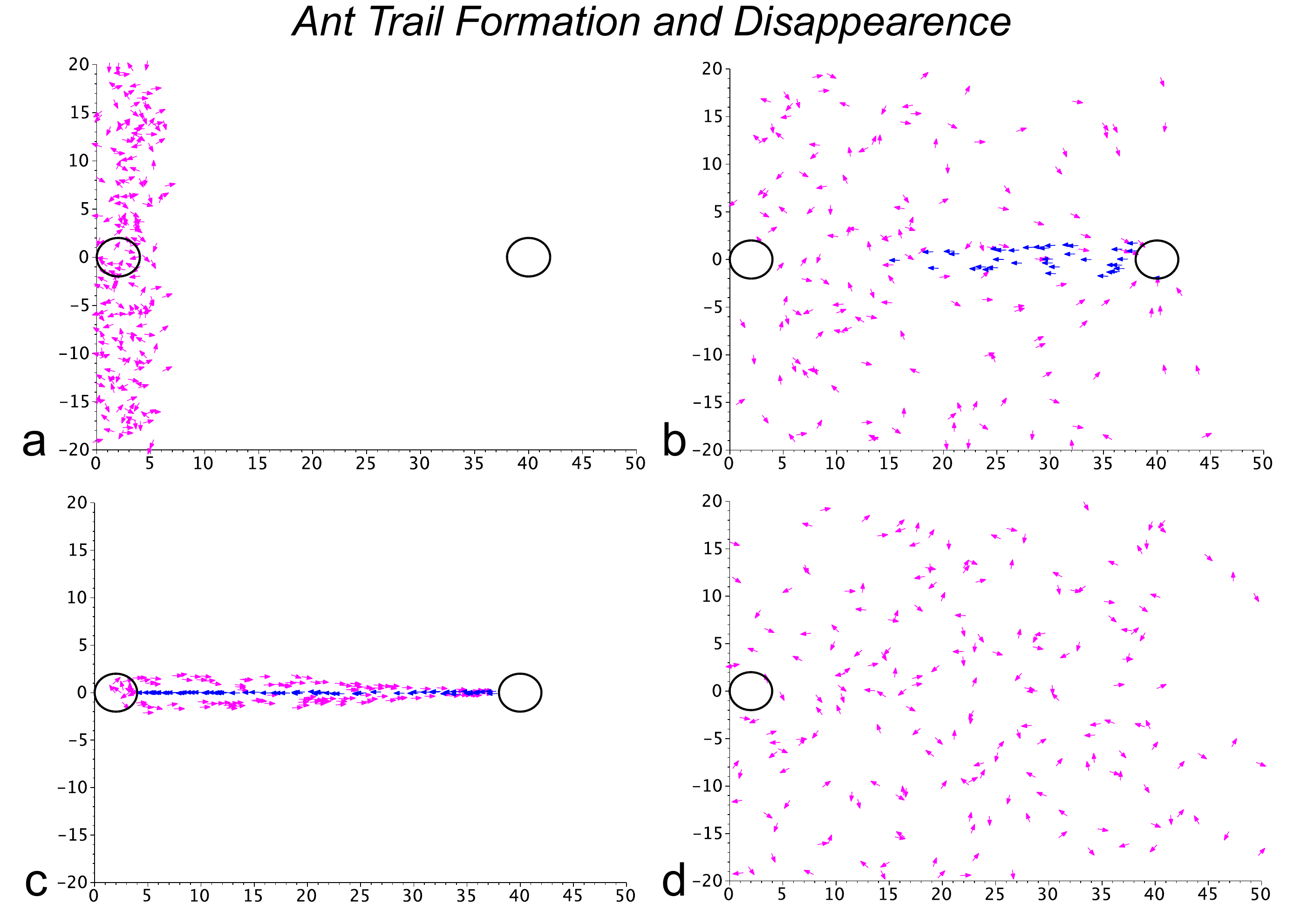}}
 \caption{Sample ant raiding simulations with foragers (purple) and returners (blue) where $N = 200$. Each arrow represents an individual ant's orientation ${\boldsymbol \omega}_i$.  The black circles denote the nest, ${\bf x}_c = (2,0)$, and food source, ${\bf x}_f = (40,0)$.  a) Initially ants are placed near the nest in non-overlapping positions with random orientation. b) Foragers begin to discover the food source and mark it with pheromone, becoming returners. c) As the pheromone diffuses more and more foragers detect the scent and begin to follow the trail to the food source. d) Once the food source is depleted the trail quickly disappears and the ants return to random foraging. See Online Resource 5.}\label{fig:app_trail}
 \end{figure}

\section{Pure pheromone diffusion model}\label{app:diff}

 As mentioned in Section~\ref{sec:disc}, one can consider a pure diffusion model for the pheromone concentration coupled with the same equations \eqref{eqn:ibm-f}-\eqref{eqn:ibm-r} governing ant dynamics.  In this setting the ants only lay pheromone at the food source the moment they become returners.  The chemical gradient is formed by diffusion of pheromone in the absence of trail laying.  We now introduce the following modified PDE for the pheromone concentration $c({\bf x}, t)$
 \begin{equation}\label{eqn:pher-new}
\begin{cases}
\displaystyle\partial_t c - \alpha\Delta c + \gamma c = \sum_{j=1}^M\delta({\bf x} - {\bf x}_f)\delta(t - t_j), &  {\bf x} \in \mathbb{R}^2, t >0\\
c({\bf x}, 0) = g({\bf x}), &  {\bf x} \in \mathbb{R}^2.
\end{cases}
\end{equation}
Here $M$ is the total number of visits before a food source is depleted, $g({\bf x})$ is a constant uniform initial distribution of pheromone, and $t_j$ is the time that the $j$th quantity of food is discovered by a forager.  This equation models each foraging ant depositing pheromone at the time of each visit.  Once the food source is depleted the pheromone concentration naturally decays to zero resulting in the trail disappearing.

This modification results in a small change in the numerical implementation. We now have an explicit analytical solution to the PDE and need to replace \eqref{eqn:fund} with
\begin{equation}
c({\bf x},t) := e^{-\gamma t}\left[\frac{1}{|V_L|} + \sum_{j=1}^M \chi_j \right], \qquad \frac{\partial c}{\partial x_i} = e^{-\gamma t}\sum_{j=1}^M \frac{\partial \chi_j}{\partial x_i}.\label{eqn:fund2}
\end{equation}
where
\begin{equation*}
\chi_j := \begin{cases}
\frac{e^{\gamma t_j}}{4\pi\alpha(t-t_j)} e^{-\frac{|{\bf x}-{\bf x}_f|^2}{4\alpha(t-t_j)}}& \quad  t > t_j,\\
0, & \quad t < t_j
\end{cases}.
\end{equation*}
By changing how the ants emit the chemical signal, we achieve an the added advantage of not requiring the computation of a time integral for each foraging ant.  One can see from Figure~\ref{fig:app_trail}, that the behavior of the ants is the same as in the prior case.  The greatest disadvantage is that this approximation to the main model presented is only valid for very short trails where diffusion of pheromone would be sufficient to attract all ants without trail laying.

 \begin{figure}
 \centerline{\includegraphics[height=1.5in]{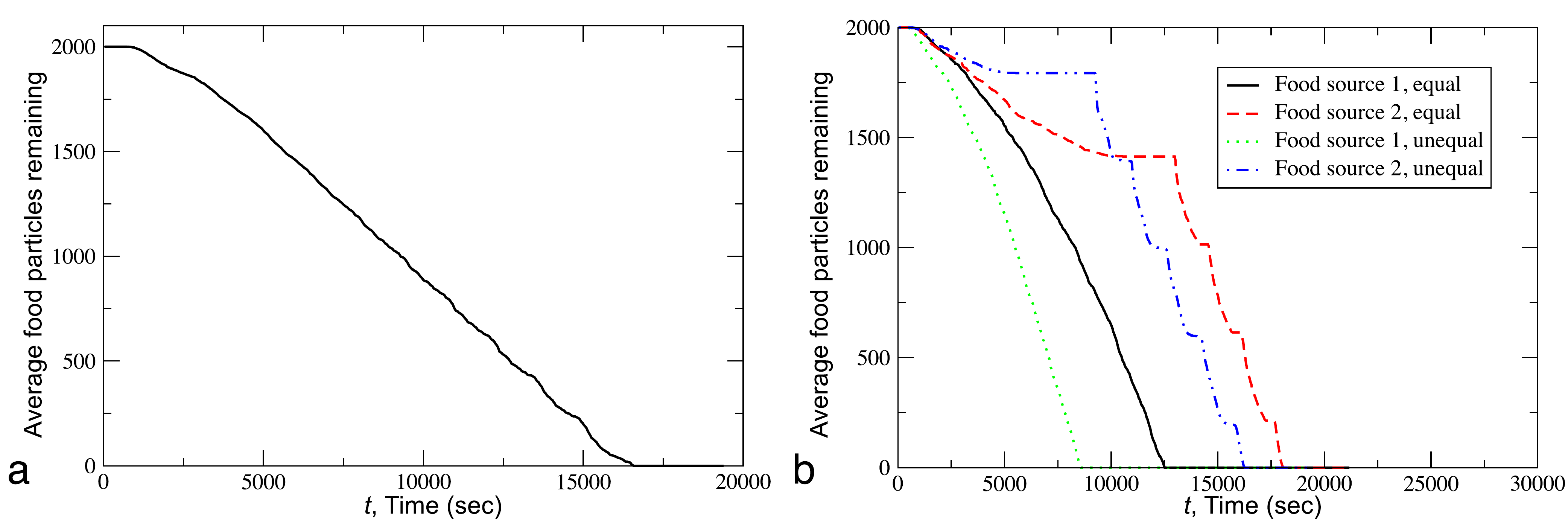}}
 \caption{Removal of food over the course of time for a) one food source or b) two food sources at equal and unequal distances from the nest.  Food depleted in 4-6 hours consistent with duration of raids from the experimental observations in \cite{Sch71,Sch40}.}\label{fig:app_food}
 \end{figure}
 
The main difference in the dynamics from the main model presented \eqref{eqn:pher}  is seen when two food sources are present.  While locally the motion of each ant may appear similar in each case, the depletion rate of each food source is different.  In the case of equidistant food sources, initially both food sources are decreasing at about the same rate, but then the depletion rate of the second food source becomes lower and eventually it is no longer visited as seen by the horizontal portion of the food count function in Figure~\ref{fig:app_food}b).  In \eqref{eqn:pher} the ants tended to all raid at the food source which was discovered first and display hardly any trail formation at the second food source until the first was depleted.  For food sources at different distances the behavior of both models is similar, the closest food source is essentially depleted first and then the second food source is visited.  Thus, ants will use all available foragers to completely deplete the closer quantity of food before moving on.  This may provide further evidence of the efficiency in which the ants seek to carry out the raiding process.

\section{Central nest location}\label{app:center}

\begin{figure}
 \centerline{\includegraphics[height=1.05in]{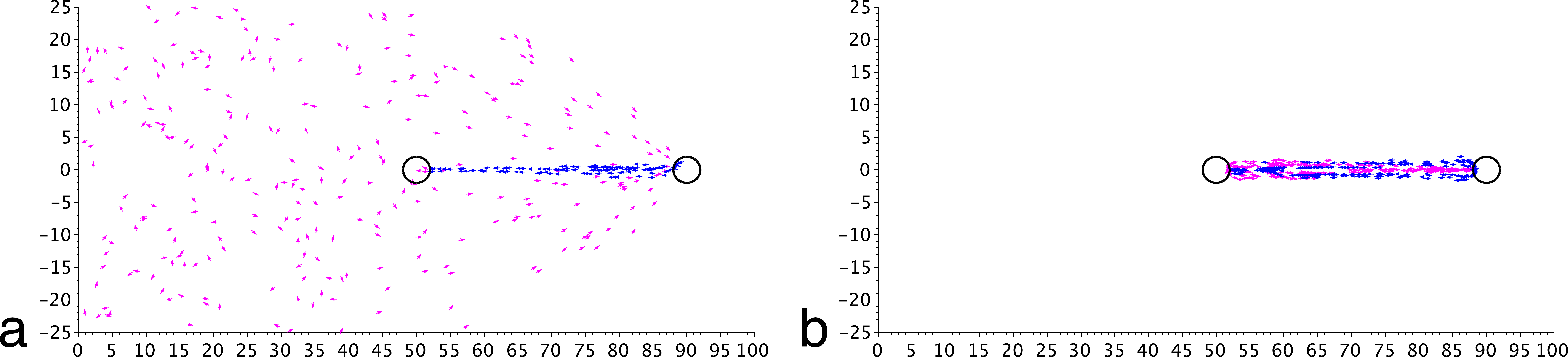}}
 \caption{Trail formation with nest in the center. Foragers (purple) and returners (blue) where $N = 400$. Each arrow represents an individual ant's orientation ${\boldsymbol \omega}_i$.  The black circles denote the nest, ${\bf x}_c = (50,0)$, and food source, ${\bf x}_f = (90,0)$. a)  Once the food source is discovered and phermone is laid along the trail, it takes a greater amount of time for it to diffuse to the ants on the opposite end of the domain.  b) Eventually all ants join in the raid resulting in a trail with the same form as in Figure~\ref{fig:trail}. See Online Resource 6.}\label{fig:center}
 \end{figure}

Some additional results are presented where the nest is located at the center of the domain.  The purpose of these images is to show that the dynamics and trail formation are essentially the same as in the scenarios presented throughout this work where the nest was closer to one edge of the domain.  Figure~\ref{fig:center} illustrates the trail formation in time.  The only difference is that with the nest in the center it takes longer to attract all the foragers, because some are now farther from any portion of the trail than the previous case.  To see the full raid please consult Online Resource 6.

\vspace{-.2in}

\end{appendix}
 
 \end{document}